\documentclass[preprint2]{proto}
\usepackage{times}
\usepackage{rotating}

\newcommand{\refs}{\par\noindent\hangindent=1pc\hangafter=1}
\begin{document}

\title{\textbf{\LARGE Identification and Dynamical Properties of Asteroid Families }}

\author {\textbf{\large David Nesvorn\'y}}
\affil{\small\em Department of Space Studies, Southwest Research Institute, Boulder}

\author {\textbf{\large Miroslav Bro\v{z}}}
\affil{\small\em Institute of Astronomy, Charles University, Prague}

\author {\textbf{\large Valerio Carruba}}
\affil{\small\em Department of Mathematics, UNESP, Guaratinguet\'a}


\begin{abstract}
\begin{list}{ } {\rightmargin 1in}
\baselineskip = 11pt
\parindent=1pc
{\small Asteroids formed in a dynamically quiescent disk but their orbits became gravitationally 
stirred enough by Jupiter to lead to high-speed collisions. As a result, many dozen large asteroids 
have been disrupted by impacts over the age of the Solar System, producing groups of fragments 
known as {\it asteroid families}. Here we explain how the asteroid families are identified,
review their current inventory, and discuss how they can be used to get insights into long-term dynamics 
of main belt asteroids. Electronic tables of the membership for 122 notable families are reported
on the Planetary Data System node. See related chapters in this volume for the significance of asteroid 
families for studies of physics of large scale collisions, collisional history of the main belt, source 
regions of the near-Earth asteroids, meteorites and dust particles, and space weathering.\\~\\~\\~}
\end{list}
\end{abstract}
\section{\textbf{INTRODUCTION}}
As witnessed by the heavily cratered surfaces imaged by spacecrafts, the chief geophysical 
process affecting asteroids is {\it impacts}. On rare occasions, the impact of a large 
projectile can be so energetic that the target asteroid is violently torn apart, and the pieces 
are thrown into space. The sites of such cosmic accidents are filled with debris that 
gravitationally accumulate into larger conglomerates, and drift away at speeds that are 
roughly commensurate with the escape speed from the original target body ($V_{\rm esc}$). 
Initially, all orbits are similar, because $V_{\rm esc} \ll V_{\rm orb}$, where $V_{\rm orb} 
\simeq 15$-20 km s$^{-1}$ is the orbital speed of main belt asteroids. On longer timescales, 
however, the orbits are altered by gravitational perturbations from planets, and the orbital 
elements of individual bodies start to diverge.

It may therefore seem challenging to identify fragments of a catastrophic collision that 
happened eons ago. Fortunately, starting with the pioneering work of K. Hirayama (Hirayama, 1918;
see also Cimrman, 1917), astronomers have developed various methods to deal with this issue (Section 2). 
Roughly speaking, these methods consist in a transformation that brings the orbital elements at the observed 
epoch to a standard, called the {\it proper elements} (Kne\v{z}evi\'c et al., 2002), that is unchanging 
in time (or, at least, would be unchanging if chaotic dynamics, non-gravitational forces, and other 
perturbations could be ignored). Thus, ideally, daughter fragments produced by breakup of a parent 
asteroid will appear as a group in space of the proper elements even gigayears after the original 
collision. These groups are called asteroid families, or {\it dynamical} families to emphasize
that they have been identified from dynamical considerations.
   
Telescopic surveys such as the Sloan Digital Sky Survey (SDSS), Wide-field Infrared Survey Explorer 
(WISE) and AKARI All-Sky Survey provide a wealth of data on physical properties of the main belt 
asteroids (Ivezi\'c et al., 2001; Mainzer et al., 2011; Usui et al., 2013). They have been used to 
cross-link the color and albedo measurements with the lists of dynamical families, in much 
the same way the spectroscopic and taxonomic data have previously been applied to this purpose 
(see Cellino et al., 2002 for a review). This work is useful to physically characterize the asteroid families 
(see chapter by Masiero et al. in this volume), including cases where two or more dynamical families overlap, 
and identify distant ``halo'' family members that would otherwise be confused with the local background 
(e.g., Bro\v{z} and Morbidelli, 2013). Given that the SDSS and WISE catalogs now contain data for over 100,000 unique 
asteroids, it has also become practical to conduct search for families in extended space, where the 
color and/or albedo data are taken into account simultaneously with the orbital elements (e.g., 
Parker et al., 2008; Masiero et al., 2013; Carruba et al., 2013a). 

\begin{figure*}[t!]
\epsscale{1.0}
\plotone{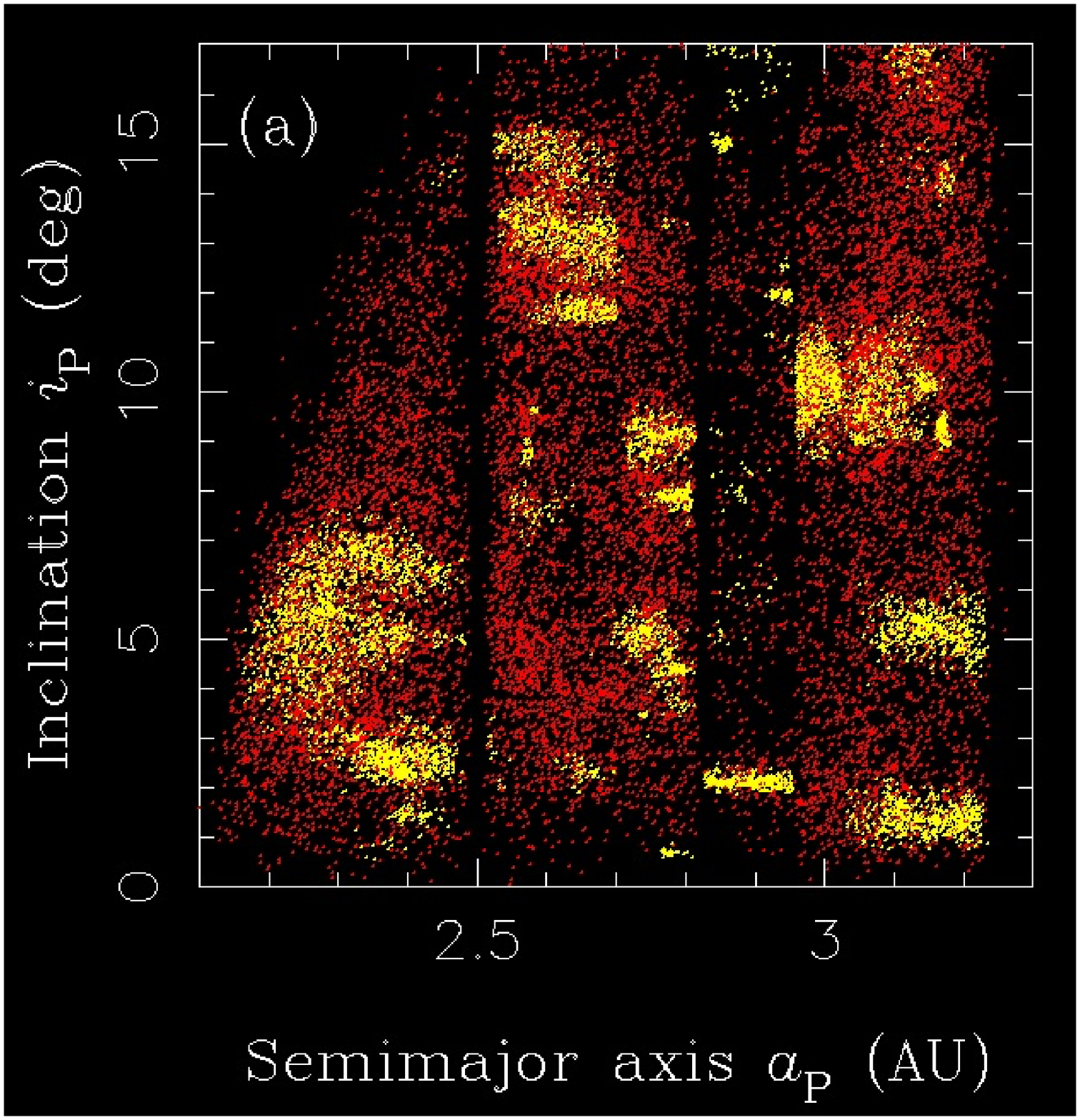}
\epsscale{0.99}
\plotone{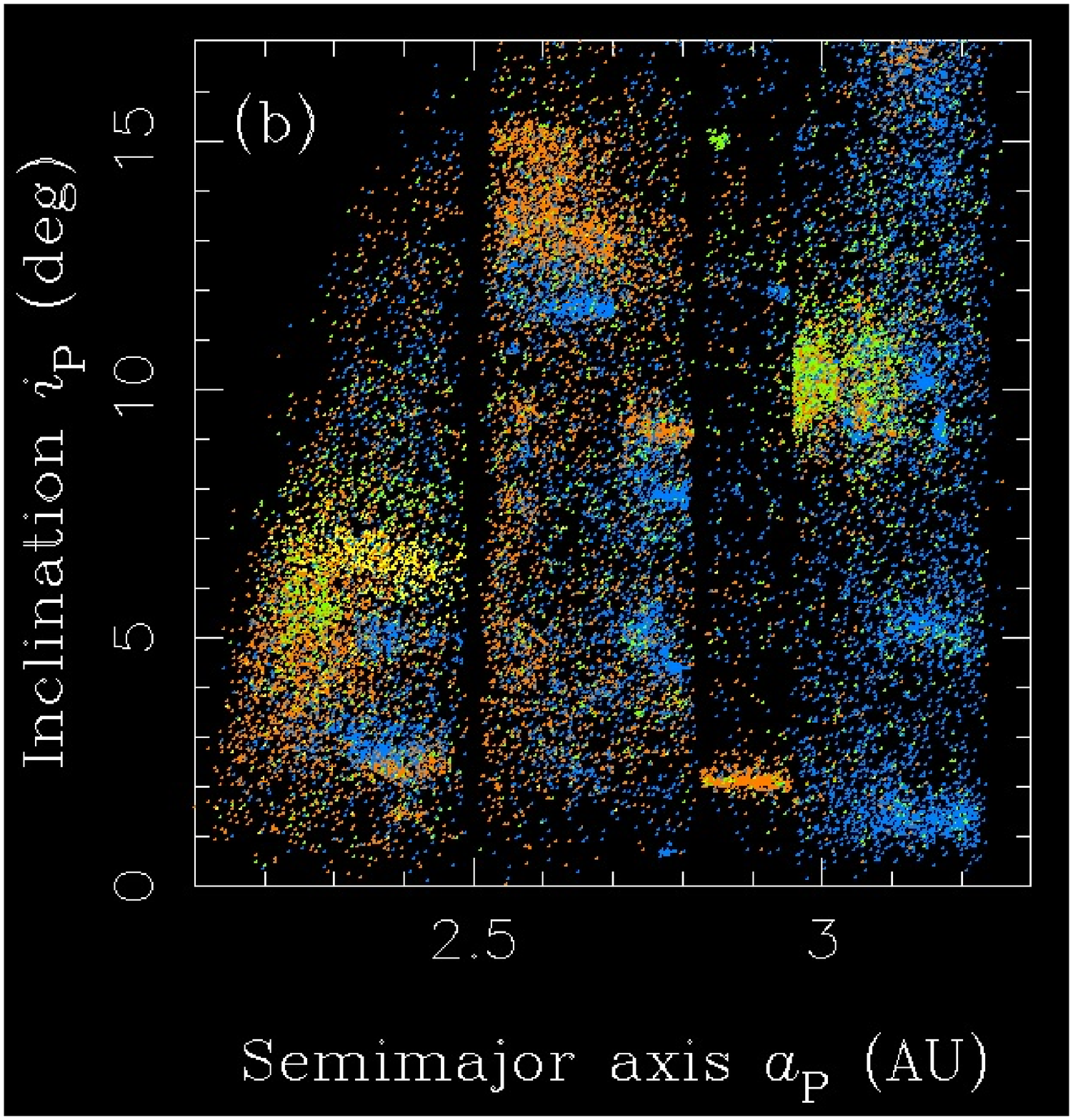}
\caption{\small (a) Clustering algorithm applied to the asteroid belt separates dynamical families 
(yellow) from the background (red). (b) Variation in reflectance properties of main belt asteroids. 
Here we plot $\simeq$25,000 asteroids that were observed by both SDSS and WISE. The color code was chosen 
to highlight the albedo/color contrast of different families.}
\label{fam}
\end{figure*}

The physical data can be used to identify {\it interlopers}. The problem of 
interlopers arises because the clustering criterion applied to identify the dynamical families is 
only a rough expression of the true membership. Unrelated asteroids that just happen to have nearby 
values of proper elements will be grouped together with the true members, and will thus appear 
in the lists of dynamical families obtained from the proper elements (e.g., Migliorini et al., 1995). 
These interlopers, especially 
the large ones, introduce ambiguity in the interpretation of impact conditions that produced individual 
families, and in the implications of these studies for asteroid interiors (cf. Michel et al., 2003;
Nesvorn\'y et al., 2006a). Here we discuss how large interlopers and true largest members in 
families can be found by applying the ``V-shape'' criterion, which is based on the notion 
that large fragments are ejected at low speeds, and have less mobility due to the Yarkovsky effect
(Section 4).

Based on the synthesis of asteroid families extracted from recent publications  (Moth\'e-Diniz et al., 2005; 
Nes\-vor\-n\'y et al., 2005; Gil-Hutton, 2006; Parker et al., 2008; Nesvorn\'y, 2010, 2012; Novakovi\'c et al., 
2011; Bro\v{z} et al., 2013; Masiero et al., 2011, 2013; Carruba et al., 2013a; Milani et al., 2014), 
we attempt to build a consensus that could serve as a starting point for future studies. 
We classify the asteroid families into {\it notable} cases (those that have a high statistical 
significance, and are thus real, and/or are notable for other reasons) and {\it candidate} families 
(less interesting cases where the statistical significance is low or cannot currently be established). 
The distinction between notable and candidate families is somewhat arbitrary, and will intentionally be 
left strictly undefined, because that is the nature of things. We expect that many candidate families 
will be confirmed with more data, and that a few notable families may fall into oblivion. 
The lists of notable families are being made available at the Planetary Data System (PDS) node, and 
are discussed in Section 7.  

A new and exciting development in the past decade was the detection of several asteroid families with 
very young formation ages. For example, the Karin family was shown to have formed only $5.8\pm0.2$~m.y. 
ago (Nesvorn\'y et al., 2002a). These cases are important, because various collisional 
and dynamical processes had little time to act on these families to alter their properties. The young 
families have thus attracted much attention from scientists studying impact physics, space weathering, 
debris disks, etc. As we explain in Section 3, the age of a young family can be determined 
by numerically integrating the orbits of its members backward in time and demonstrating that they
converge to each other at some specific time in the past. This is the time of a breakup, and the family 
age, $t_{\rm age}$, is the time elapsed from the breakup event. 

The method of backward integration of orbits only works for the families with $t_{\rm age} \lesssim 10$ m.y. 
This is because dynamics of main-belt asteroids on longer timescales is governed by chaos, 
encounters with (1) Ceres and other large asteroids, and non-gravitational forces. A complementary 
statistical method for the estimation of family age has been developed in Vokrouhlick\'y et al. (2006a,b). The 
method tracks, in detail, how the family structure in semimajor axis changes over time as the family 
members drift away by the Yarkovsky effect (Section 5). The semimajor axis spread 
of an older family will generally be greater than that of a younger family. A~compilation of formation ages 
of the asteroid families can be used to constrain how the population of the asteroid belt collisionally 
evolved over time (e.g., Bottke et al., 2005a,b; Cibulkov\'a et al., 2014), and how asteroid surfaces age 
by space weathering (chapter by Brunetto et al. in this volume). 
\section{\textbf{IDENTIFICATION METHOD}}
Here we discuss the standard method to identify asteroid families. This method consists of the (1) computation 
of proper elements, or other elements unchanging with time, for asteroids with well-known orbits, (2) identification of 
concentrations or groups of asteroids in proper element space, and (3) establishing the statistical 
significance of identified groups. These steps are discussed in Sections 2.1, 2.2 and 2.3.
In Section 2.4, we examine the ``overlap problem'' where two or more families overlap in proper 
element space, and need to be separated. For this it is useful to consider families in 
{\it extended} space with physical data being included in addition to the proper elements.
The search in extended space can also lead to the identification of new families (Section 2.5).
Very young families ($t_{\rm age}\lesssim 1$ m.y.), for which the member orbits have not had time 
to differentially precess away from each other, can also be identified as groups in space of 
the {\it osculating} orbital elements (Section~2.6).  
   
\bigskip
\noindent
\textbf{2.1 PROPER ELEMENTS}
\bigskip

The ejection speeds of sizable fragments produced by collisional breakups of main belt asteroids are 
generally much smaller than their orbital speeds. The fragments will therefore initially cluster near 
the original orbit of their parent body, and will appear as such if the subsequent effects of planetary 
perturbations are removed by projecting orbits into space of proper elements. The three most 
useful proper elements are: the proper semimajor axis ($a_{\rm P}$), the proper eccentricity ($e_{\rm P}$), 
and the proper inclination ($i_{\rm P}$). They are close equivalents of their osculating element 
counterparts in that they define the size, elongation and tilt of orbits (see Note 1 in Section 9).

The definition of proper elements as quasi-integrals of asteroid motion, and the methods 
used to compute them, were explained in the Asteroids III book (Kne\v zevi\'c et al., 
2002). As these definitions and methods have not changed much, we do not discuss them here in detail. 
In brief, the proper elements are obtained from the instantaneous osculating orbital elements by 
removing periodic oscillations produced by gravitational perturbations of planets. This can be 
done analytically, using perturbation theory (Milani and Kne\v{z}evi\'c, 1990, 1994), or numerically, 
by integrating the orbits and applying the Fourier analysis (Kne\v{z}evi\'c and Milani, 2000; 
Kne\v{z}evi\'c et al., 2002) (Note 2).


The computation of analytic proper elements is relatively CPU inexpensive.
They are made publicly available by A. Milani for both numbered and unnumbered multi-opposition 
asteroids at the AstDyS node (Note 3). The analytic proper 
elements lose precision for highly-inclined orbits, because the expansion of the gravitational 
potential used to calculate them has poor convergence for high inclinations. The more-precise 
synthetic proper elements (precision generally at least 3 times better than that of analytic 
elements), on the other hand, require a much larger CPU investment, and are only made available 
at the AstDys node for the numbered asteroids.


\bigskip
\noindent
\textbf{2.2 CLUSTERING ALGORITHM}
\bigskip

When these methods are applied to the asteroid belt, things are brought into focus with dozens of 
obvious clumps, asteroid families, emerging from the background (Figure \ref{fam}a). 
To identify an asteroid family, researchers apply a clustering algorithm to the distribution 
of asteroids in ($a_{\rm P},e_{\rm P},i_{\rm P}$) space. The most commonly used algorithm is 
the Hierarchical Clustering Method (HCM; Zappal\`a et al., 1990) (Note 4), which defines a cutoff distance, 
$d_{\rm cut}$, and requires that the length of the link between two neighboring orbits clustered 
by the algorithm is $d = d(a_{\rm P},e_{\rm P},i_{\rm P})< d_{\rm cut}$. A common definition of 
distance is $d^2 \equiv (n a_{\rm P})^2 (k_a (\delta a_{\rm P}/a_{\rm P})^2 + k_e (\delta e_{\rm P})^2 + 
k_i (\delta i_{\rm P})^2)$, where $n$ is the orbital frequency, ($\delta a_{\rm P},\delta e_{\rm P},
\delta \sin i_{\rm P}$) is the separation vector between orbits in 3D space of proper elements, and 
$(k_a$,$k_e$,$k_i)$ are coefficients of the order of unity.

The main advantage of the HCM over other methods is that there is no strong assumption built into the 
HCM about the shape of an asteroid family in proper element space. This is because the chain created by 
linking nearby orbits can track down family members even if their orbits dynamically evolved 
to produce an unusual overall shape. A prime example of this is the case of the Koronis family 
which is split into two parts by the secular resonance $g+2g_5-3g_6=0$ at $\simeq$2.92~AU, where $g$
is the apsidal frequency of an asteroid, and $g_5$ and $g_6$ and the 5th and 6th apsidal 
frequencies of the planetary system. The part of the Koronis family with $a_{\rm P}>2.92$~AU has larger 
eccentricity than the part with $a_{\rm P}<2.92$~AU, because family members drifting by the Yarkovsky effect
from $a_{\rm P}<2.92$~AU have their eccentricities increased by interacting with the $g+2g_5-3g_6=0$ resonance
(Bottke et al., 2001). 

The main disadvantage of the standard HCM, which becomes increasingly difficult to overcome with 
inclusion of numerous small asteroids in the new catalogs, is the problem of {\it chaining}. This problem
arises because small fragments are typically ejected at higher speeds and have larger mobility
due to the Yarkovsky effect. They therefore spread more, tend to be distributed 
more homogeneously throughout the main belt, and create bridges between different families if
a single (large) value of $d_{\rm cut}$ is used. Clearly, $d_{\rm cut}$ should be set proportional 
to the asteroid size, or inversely proportional to the absolute magnitude~$H$. Expressing this dependence, 
however, adds additional parameters to the HCM and makes the whole identification procedure more 
complex. Therefore, in reality, it is preferred to bypass the problem of chaining by artificial means (e.g., 
cuts in proper element space applied to deal with individual cases), or the proportionality is 
approximated by a two-step method with different cutoffs for small and large bodies (Milani et al., 2014).     
   
A tricky part of the HCM algorithm is the choice of the cutoff distance. If the value of $d_{\rm cut}$ 
is too small, many dispersed but real families will remain unnoticed and large families will artificially 
be split into parts. If the value is too large, the algorithm will clump different families together, and 
will identify irrelevant clumps produced by random fluctuations. While many asteroid families can be identified 
for a wide range of $d_{\rm cut}$ values, and are real beyond doubt, some cases require a specific choice
of $d_{\rm cut}$ and can potentially be confused with random fluctuations. Clearly, the {\it statistical 
significance} of the identified groups, or their insignificance, needs to be established before 
proceeding further. 

\bigskip
\noindent
\textbf{2.3 STATISTICAL SIGNIFICANCE}
\bigskip

To make reasonably sure that families identified from the HCM are real one can opt for a conservative 
choice of $d_{\rm cut}$. This can be done, for example, by collecting asteroids in a given region 
of proper element space and redistributing them randomly in that region. The HCM applied to this
artificial distribution will reveal that the largest identified group with given $d_{\rm cut}$ contains 
$N^*(d_{\rm cut})$ members. Now, this procedure can be repeated, say, one thousand times, recording 
the largest $N^*(d_{\rm cut})$ obtained from these trials. We can then be 99\% confident that random 
fluctuations cannot produce groups with more that $N^*(d_{\rm cut})$ members (this conservative 
estimate includes a 99\% confidence interval computed by the Wilson score interval approximation).
Any group identified in the {\it real} distribution with $N>N^*(d_{\rm cut})$ members is therefore 
reasonably likely to be real. Higher confidence levels can be achieved by increasing the sample size. 

This basic concept, and various modifications of it, is known as the Quasi Random Level (QRL; 
Zappal\`a et al., 1994). In the ideal world, the QRL would be the ultimate solution to the family 
identification problem: just choose $d_{\rm cut}$ and pick up all clumps with more than  $N>N^*(d_{\rm cut})$; 
those clumps are real. Then, there is the real world. First, the number density in proper element space is 
variable due to the primordial sculpting of the main belt and resonances (e.g., Minton and Malhotra, 2009). 
Applying a {\it global} QRL value in (parts of) the main belt may therefore lead to unsatisfactory results. Second, 
families do not live in isolation but are frequently close to each other, overlap, and/or are surrounded 
by empty regions. This introduces an ambiguity in the QRL definition, because it is not clear a priori 
what region in ($a_{\rm P},e_{\rm P},i_{\rm P}$) space should be considered to define the {\it local} 
QRL in the first place. Results may depend on this choice.

In practice, the first choice made is often the minimum number of group members, $N_{\rm min}$,
that is considered to be interesting. Then, the cutoff distance $d_{\rm cut}$ in 
some local region in ($a_{\rm P},e_{\rm P},i_{\rm P}$) is defined such that groups of $N>N_{\rm min}$
members cannot be produced with $d_{\rm cut}$ by random fluctuations. All groups
with $N>N_{\rm min}$ are then treated as meaningful asteroid families. Different researchers made
different choices: $N_{\rm min}=5$ in Zappal\`a et al. (1990), $N_{\rm min}=100$ in Parker et al. 
(2008), and $N_{\rm min}=10$-20 in most other publications. The two disadvantages of this method 
are that: meaningful asteroid families with $N<N_{\rm min}$ members are explicitly avoided, and 
$d_{\rm cut}(N_{\rm min})$ depends on the population density in proper element space and must be 
recomputed when a new classification is attempted from ever-growing catalogs. 

\begin{figure}[t!]
\epsscale{1.0}
\plotone{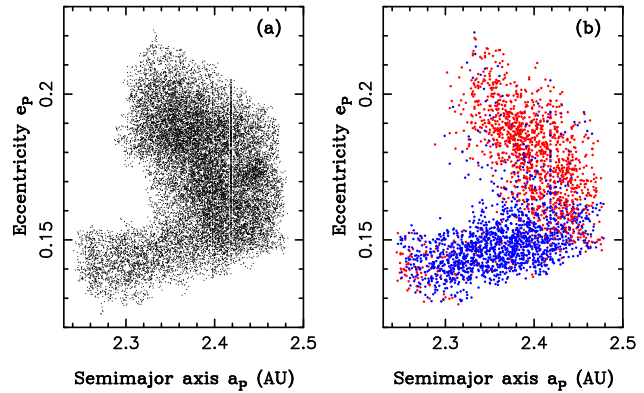}
\caption{\small The Nysa-Polana complex. (a) The HCM applied to this region of the inner main belt reveals 
a major concentration of asteroids with $2.25<a_{\rm P}<2.48$ AU and $0.13<e_{\rm P}<0.22$. The shape 
of the concentration in the $(a_{\rm P},e_{\rm P})$ projection is unusual and difficult to interpret. 
(b) The WISE albedos of members of the Nysa-Polana complex: black for $p_V < 0.15$ and gray for $p_V > 0.15$. 
It becomes clear with the albedo information that the Nysa-Polana complex is two overlapping groups with 
distinct albedos. Furthermore, based on the V-shape criterion (Section 4), the low-albedo group is
found to consist of two asteroid families (the Polana and Eulalia families; Walsh et al., 2013). The 
vertical feature at $a_{\rm P}\simeq2.42$ AU is the 1:2 mean motion resonance with Mars.}
\label{nysa}
\end{figure}

Another approach to this problem is to identify all groups, even if they have only a few members,
and establish their statistical significance {\it a posteriori}. Those that are judged to be 
insignificant are subsequently discarded and do not appear in the final lists. To determine the 
statistical significance of a group, one can generate mock distributions and apply the HCM to them. 
For example, the high statistical significance of the Karin family, which is embedded in the much 
larger Koronis family, can be demonstrated by generating thousand orbital distributions corresponding 
to the Koronis family, and applying the HCM to each one (Nesvorn\'y et al., 2002a). With $d_{\rm cut} = 
10$ m s$^{-1}$, no concentrations in this input can be found containing more than a few dozen members, 
while the Karin family currently has 541 known members. Therefore, the Karin family is significant 
at a greater than the 99\% level (again including a 99\% confidence interval of the estimate). A systematic 
application of this or similar statistical arguments can be quite laborious if many borderline cases 
need to be resolved.  

\bigskip
\noindent
\textbf{2.4 OVERLAP PROBLEM AND INTERLOPERS}
\bigskip

The overlap between different families has become more of a problem with a progressively 
larger share being taken in the proper element catalogs by small, km and sub-km size asteroids. 
This is because small fragments are generally launched at higher speeds, and are therefore initially 
spread in a larger volume in ($a_{\rm P},e_{\rm P},i_{\rm P}$) space. Mainly, however, the problem is 
caused by the larger mobility of small fragments due to the Yarkovsky effect. For example, the mean 
drift rate of a diameter $D=1$ km main belt asteroid is estimated to be $\simeq10^{-4}$ AU/m.y. 
(Bottke et al., 2006). The km-size members of a 1-g.y. old family are therefore expected to be dispersed over 
$\simeq 0.2$ AU (the additional factor of two accounts here for fragments having different spin 
orientations, and thus ${\rm d}a/{\rm d}t<0$ or ${\rm d}a/{\rm d}t>0$), which is roughly 1/5 of the extension of 
the whole main belt. In addition, drifting asteroids encounter orbital resonances and can be 
dispersed by them in $e_{\rm P}$ and $i_{\rm P}$ as well.       
 
A good illustration of this is the case of the Flora and Vesta families in the inner main belt. 
To separate these families from each other down to their smallest members, the scope of the HCM can be 
restricted by an artificial cut in proper element space. Alternatively, one can first apply the HCM 
to the distribution of large members, thus identifying the core of each family, and then proceeding 
by trying to ``attach'' the small members to the core. This second step must use a lower $d_{\rm cut}$ 
value than the first step to account for the denser population of smaller asteroids.
In practice, this has been done by applying an absolute magnitude cutoff, $H^*$, with $H<H^*$ for the core 
and $H>H^*$ for the rest. In the low-$i$ portion of the inner main belt, where the Flora and Vesta families 
reside, Milani et al. (2014) opted to use $H^*=15$, and identified cores of families with $N_{\rm min}=17$ and 
$d_{\rm cut}=60$ m s$^{-1}$, and small members with $N_{\rm min}=42$ and $d_{\rm cut}=40$ m s$^{-1}$. 

Another solution to the overlap problem is to consider the {\it physical} properties of asteroids. Previously,
the spectroscopic observations of members of dynamical families have been used to: (1) establish the
physical homogeneity of asteroid families (the difference between physical properties of members of the 
same family tends to be smaller than the differences between physical properties of different families), 
and (2) identify large interlopers (asteroids classified as family members based on proper elements but 
having spectroscopic properties distinct from the bulk of the family).
With the color and albedo data 
from the SDSS and WISE (Note 5), the physical homogeneity of asteroid families has been demonstrated to hold down 
to the smallest observable members (Ivezi\'c et al., 2001; Parker et al., 2008; see Figure \ref{fam}b). A straightforward 
implication of this result is that the interior of each disrupted body was (relatively) homogeneous, at least on a 
scale comparable to the size of the observed fragments ($\sim$1-100 km) (Note 6).

The physical homogeneity of asteroid families can be used to identify {\it interlopers} as those members 
of a dynamical family that have color and/or albedo significantly distinct from the rest of the family. 
The number density of apparent color/albedo interlopers in a family can then be compared with the number 
density of the same color/albedo asteroids in the immediate neighborhood of the family. Similar densities 
are expected if the identified bodies are actual interlopers in the family. 
If, on the other hand, the density of color/albedo outliers in the family is found to be 
substantially higher than in the background, this may help to rule out the interloper premise, and 
instead indicate that: (i) the disrupted parent body may have been heterogeneous, or (ii) we are looking at 
two or more overlapping dynamical families with distinct color/albedo properties. Finally, as for (ii), 
it is useful to verify whether the family members with different color/albedo properties also have  
different proper element distributions, as expected if breakups happened in two (slightly) different 
locations in proper element space (e.g., the Nysa-Polana complex; see Figure \ref{nysa}). 

\begin{figure*}[t!]
\epsscale{0.71}
\plotone{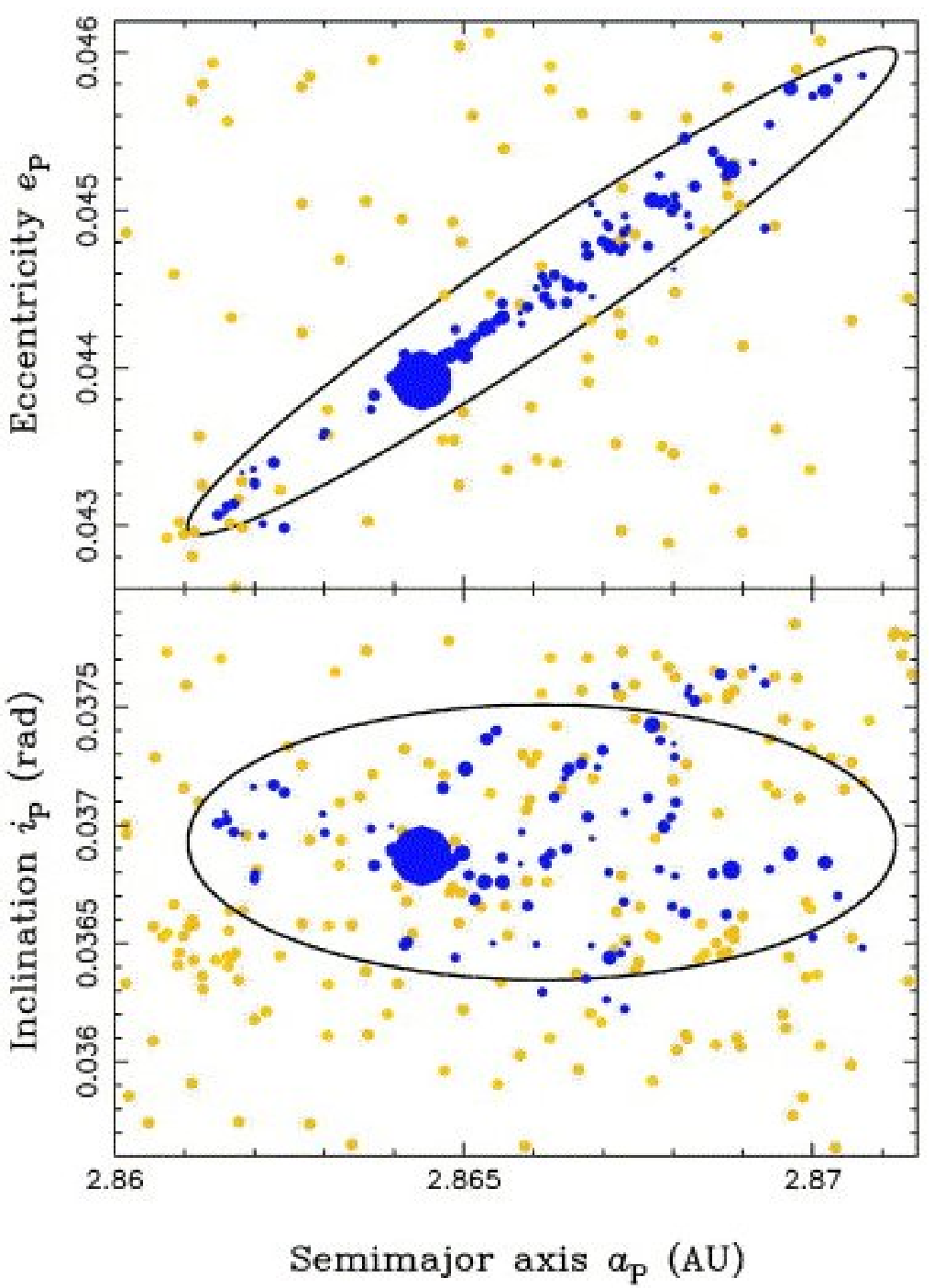}
\epsscale{1.0}
\plotone{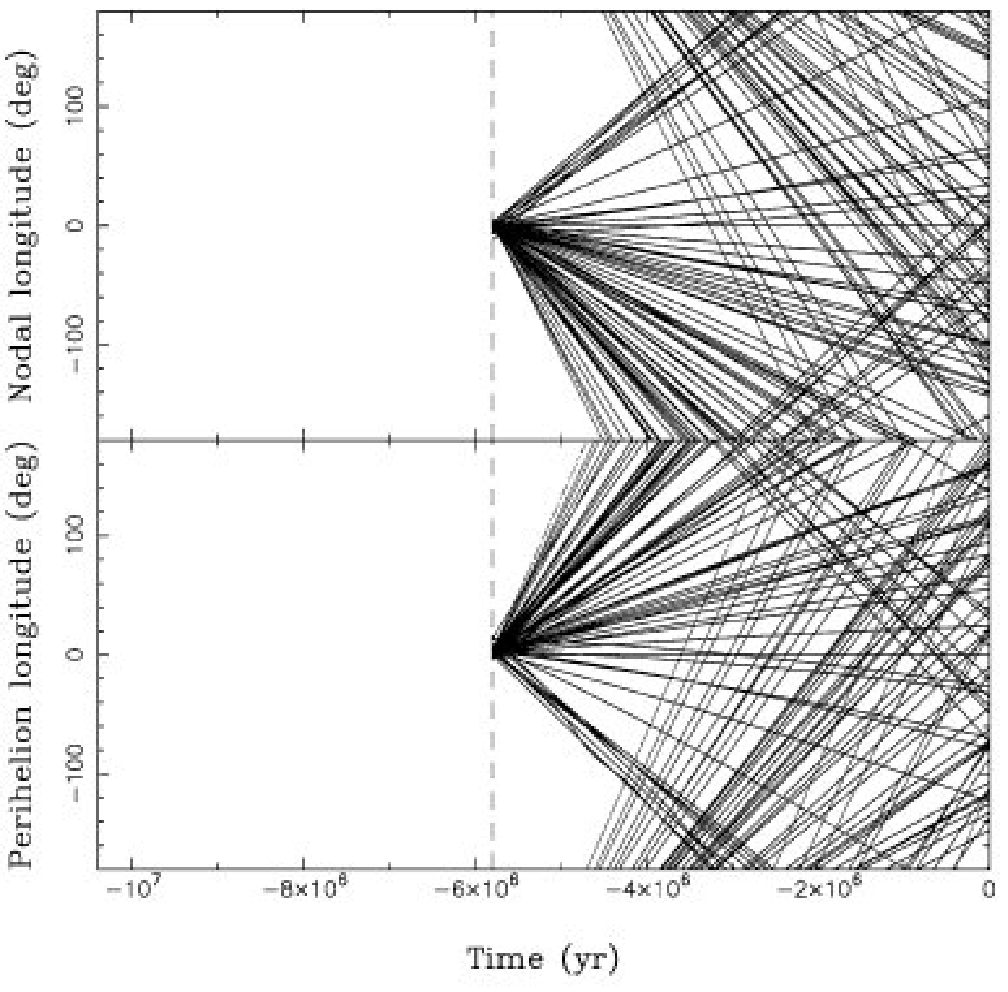}
\caption{\small 
{\bf Left panels}: Proper elements of members of the Karin family. The size of each dark 
symbol is proportional to the diameter of a family member. Light gray dots indicate background bodies 
near the Koronis family. The black ellipses show the proper orbital elements of test bodies launched at
$15$ m s$^{-1}$ from $a_{\rm P} =  2.8661$ AU, $e_{\rm P} = 0.04449$ and $i_{\rm P}=0.03692$, assuming 
that $f = 30^\circ$ and $\omega+f = 45^\circ$, where $f$ and $\omega$ are the true anomaly and 
perihelion argument of the disrupted body at the time of the family-forming collision.
{\bf Right panels}: The convergence of angles at $5.8$ m.y. ago demonstrates that the Karin family 
was created by a parent asteroid breakup at that time. The plot shows past orbital histories of ninety 
members of the Karin family: (top) the proper nodal longitude, and (bottom) the proper perihelion longitude. 
Values of these angles relative to (832) Karin are shown. At $t=−5.8$ m.y. (broken vertical line), 
the nodal longitudes and perihelion arguments of all ninety asteroids become nearly the same, as 
expected if these bodies had initially nearly the same orbits. Adapted from Nesvorn\'y and
Bottke (2004).}   
\label{karin}
\end{figure*}

\bigskip
\noindent
\textbf{2.5 FAMILIES IN EXTENDED SPACE}
\bigskip

Another useful strategy is to include the color and/or albedo information {\it directly} 
in the clustering algorithm. This can be done by first separating the main belt into two (or more) 
populations according to their color and albedo properties. For example, asteroids in the S-complex can 
be separated from those in the C/X-complex based on the SDSS colors (Nesvorn\'y et al. 2005), and
the high-albedo asteroids can be separated from the low-albedo asteroids based on the albedo 
measurements of WISE (Masiero et al., 2013). The HCM is then applied to these populations separately. This method 
is capable of identifying small/dispersed S-complex families in the C/X-type background, and vice-versa, or 
low-albedo families in the high-albedo background, and vice-versa. It can also be useful to characterize 
the so-called family ``halos'' (Section 6.4). 

A more general method for including the color/albedo information in the clustering algorithm consists 
in the application of the HCM in space of increased dimension (e.g., Parker et al. 2008; Carruba et al., 2013a). 
When considering the proper elements and SDSS colors, the distance in 5D can be defined as $d_2^2 \equiv d^2 + 
n^2 a_{\rm P}^2 (k_1 (\delta C_1)^2 + k_2 (\delta C_2)^2)$, where $d$ is the distance in 3D space of proper 
elements defined in Section 2.2, $C_1$ and $C_2$ are two diagnostic colors defined from the SDSS
(Ivezi\'c et al., 2001; Nesvorn\'y et al., 2005), and $k_1$ and $k_2$ are coefficients whose magnitude is 
set to provide a good balance between the orbital and color dimensions (e.g., Nesvorn\'y et al., 2006b). 
Similarly, we can define $d_3^2 \equiv d^2 + n^2 a_{\rm P}^2 k_p (\delta p_V)^2$ (in 4D) and $d_4^2 \equiv d_2^2 + n^2 a_{\rm P}^2 k_p 
(\delta p_V)^2$ (in 6D) to include the measurements of albedo $p_V$ from WISE. The $d_4$ metric 
applies the strictest criteria on the family membership, because it requires that the family members have
similar proper elements, similar colors, and similar albedos. Note, however, that this metric can only 
be applied to a reduced set of main belt asteroids for which the proper elements, colors and albedos 
are {\it simultaneously} available (presently $\simeq$25,000; Figure \ref{fam}b). 

\bigskip
\noindent
\textbf{2.6 VERY YOUNG FAMILIES IN ORBITAL ELEMENT SPACE}
\bigskip

Short after family's creation, when the mutual gravity effects among individual fragments cease to 
be important, the fragments will separate from each other and find themselves moving on heliocentric 
orbits. Initially, they will have very tightly clustered orbits with nearly the same values of the 
osculating orbital angles $\Omega$, $\varpi$ and $\lambda$, where $\Omega$ is the nodal longitude,
$\varpi$ is the apsidal longitude, and $\lambda$ is the mean longitude. The debris cloud will be subsequently 
dispersed by the (i) {\it Keplerian shear} (different fragments are ejected with different velocity vectors, 
have slightly different values of the semimajor axis, and therefore different orbital periods) 
and (ii) {\it differential precession} of orbits produced by planetary perturbations. 

As for (i), the fragments will become fully dispersed along an orbit on a timescale 
$T_n = \pi/(a \partial n / \partial a) (V_{\rm orb}/\delta V)\\=(P/3)(V_{\rm orb}/\delta V)$, where $P=2$-4 
yr is the orbital period and $\delta V$ is the ejection speed. With $\delta V=1$-100 m s$^{-1}$, 
this gives $T_n=300$-30,000 yr. Therefore, the dispersal of fragments along the orbit is relatively 
fast, and the clustering in $\lambda$ is {\it not} expected if a family is older than a few tens of 
thousand years. 

The dispersal of $\Omega$ and $\varpi$ occurs on a time scale
$T_f = \pi/(a \partial f / \partial a) (V_{\rm orb}/\delta V)$, where the frequency $f=s$ or $g$.
For example, $\partial s / \partial a=-70$ arcsec yr$^{-1}$ AU$^{-1}$ and $\partial g / \partial 
a=94$ arcsec yr$^{-1}$ AU$^{-1}$ for the Karin family ($a \simeq 2.865$~AU). With 
$\delta V=15$ m s$^{-1}$ (Nesvorn\'y et al., 2006a) and $V_{\rm orb}=17.7$ km s$^{-1}$, this gives $T_s=3.8$ m.y. and 
$T_g=2.8$~m.y. Since $t_{\rm age} > T_s$ and $t_{\rm age} > T_g$ in this case, the distribution 
of $\Omega$ and $\varpi$ for the Karin family is not expected to be clustered at the present 
time (Figure \ref{karin}). Conversely, the clustering of $\Omega$ and $\varpi$ would be expected 
for families with $t_{\rm age} \lesssim 1$ m.y.     

This expectation leads to the possibility that the families with $t_{\rm age} \lesssim 1$ m.y. could be 
detected in the catalogs of {\it osculating} orbital elements (Marsden, 1980; Bowell et al., 1994), where
they should show up as clusters in 5D space of $a$, $e$, $i$, $\varpi$ and $\Omega$. The search in 
5D space of the osculating orbital elements can be performed with the HCM method and metric 
$d_5^2 = d^2 + (na)^2 (k_\Omega (\delta \Omega) ^2 + k_\varpi (\delta \varpi)^2)$, where $d=d(a,e,i)$
was defined in Section 2.2, and $k_\Omega$ and $k_\varpi$ new coefficients. [Different metric functions 
were studied by Ro\v{z}ek et al. (2011), who also pointed out that using the {\it mean} elements, 
instead of the osculating ones, can lead to more reliable results.] 

This method was first successfully used in practice for the identification of the Datura family (Nesvorn\'y 
et al., 2006c), and soon after for the discovery of the asteroid {\it pairs} (Vokrouhlick\'y and Nesvorn\'y, 
2008). The Datura family now consists of 15 known members ranging in size from $\simeq$10-km-diameter 
object (1270) Datura to sub-km fragments. They have $\Omega$ and $\varpi$ clustered to within a few 
degrees near 98$^\circ$ and 357$^\circ$, respectively. The age of the Datura family is only $530 \pm 
20$ k.y., as estimated from the backward integration of orbits (Vokrouhlick\'y et al., 2009). Table 1 
reports other notable cases of families with $t_{\rm age}<1$ m.y.

\begin{figure}[t!]
\epsscale{1.0}
\plotone{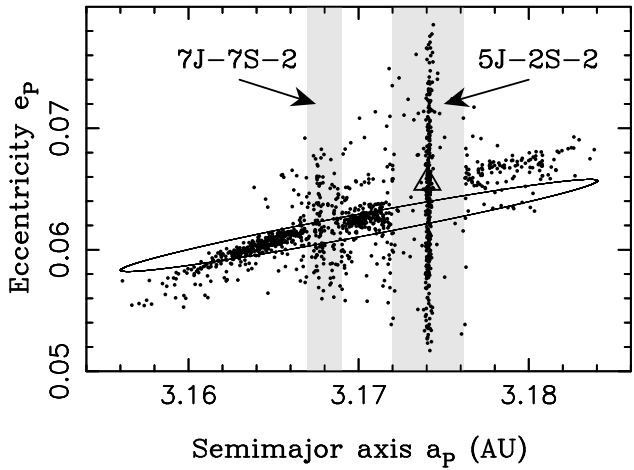}
\caption{\small The Veritas family. Here we plot the proper elements of 1294 members of the Veritas 
family identified from a new catalog (August 2014). (490) Veritas itself is marked by a triangle.
The gray vertical strips indicate two three-body resonances which act to diffuse the orbits of family 
members in $e_{\rm p}$. The black ellipse shows the orbital elements of test bodies launched at $\delta V 
= 35$~m~s$^{-1}$ from $a_{\rm P} =  3.17$~AU and $e_{\rm P} = 0.0062$, assuming that $f = 30^\circ$, where $f$ 
is the true anomaly of the disrupted body at the time of the family-forming collision.}
\label{veritas}
\end{figure}

\section{\textbf{DETECTION OF RECENT BREAKUPS}}
The detection of families with very young formation ages was one of the highlights of asteroid research 
in the past decade. A poster child of this exciting development is the Karin family, part of the 
larger Koronis family, that was shown to have formed only $5.8\pm0.2$ m.y. ago (Nesvorn\'y et al., 2002a).
The Karin family was identified by the traditional means, using the HCM on proper elements. The 
diagonal shape of this family in the $(a_{\rm P},e_{\rm P})$ projection is a 
telltale signature of a recent breakup, because this initial shape is expected if a breakup occurs near 
the perihelion of the parent body orbit (Figure \ref{karin}). In this case, the tilt $\alpha$ of the family shape in the 
$(a_{\rm P},e_{\rm P})$ projection is expected to be, and indeed {\it is} in the case of the Karin family, 
$\tan \alpha = \Delta e_{\rm P} / (\Delta a_{\rm P}/a_{\rm P}) = (1-e_{\rm P}) = 0.956$. Thus, $\alpha 
\simeq 45^{\circ}$. All other known families, with an exception of the similarly-young Veritas family, 
have $\alpha \sim 0$ (i.e., are nearly horizontal features in $a_{\rm P},e_{\rm P}$). This is because 
these families are old and their original shape was stretched in $a_{\rm P}$ by the Yarkovsky effect 
(e.g., Dell'Oro et al., 2004). 

\begin{table*}[t]
\caption{Recently formed asteroid families.}
\begin{center}
\begin{tabular}{lrll}
\hline
Family/Pair            & $t_{\rm age}$       & References     & Notes \\
\hline
(832) Karin            & $5.75\pm0.05$ m.y.  & Ne02,NB04      & 2.1$^\circ$ dust band \\
(158) Koronis(2)       & 10-15 m.y.          & MH09               & near (832) Karin \\
(490) Veritas          & $8.3\pm0.5$ m.y.    & Ne03,F06,T07   & 9.3$^\circ$ band, late Miocene dust shower \\ 
(656) Beagle           & $\sim$10 m.y.       & Ne08           & 1.4$^\circ$ band, member Elst-Pizzaro? \\
(778) Theobalda        & $6.9\pm2.3$ m.y.    & No10           & $t_{\rm age}$ needs to be confirmed\\
(1270) Datura          & $530\pm20$ k.y.     & Ne06,V09       & identified in 5D, E/F dust band? \\  
(2384) Schulhof        & $780\pm100$ k.y.    & VN11           & secondary breakup event?\\   
(4652) Iannini         & $\lesssim5$ m.y.    & Ne03,W08       & chaotic dynamics\\
(5438) Lorre           & $1.9\pm0.3$ m.y.    & No12a          & $i_{\rm P} \simeq 28^\circ$  \\
(14627) Emilkowalski   & $220\pm30$ k.y.     & NV06           & only 3 members known\\      
(16598) 1992 YC2       & 50-250 k.y.         & NV06           & only 3 members known\\
(21509) Lucascavin     & 300-800 k.y.        & NV06           & only 3 members known\\
(300163) P/2006 VW139  & $7.5\pm0.3$ m.y.    & No12b          & main belt comet \\
P/2012 F5 (Gibbs)      & $1.5\pm0.1$ m.y.    & No14           & main belt comet \\
\hline
\label{youngfam}
\end{tabular}
\end{center}
{\small {\bf References:} 
Ne0X = Nesvorn\'y et al. (200X), NoXX = Novakovi\'c et al. (20XX), VNXX = Vokrouhlick\'y and 
Nesvorn\'y (20XX), VXX = Vokrouhlick\'y et al. (20XX), NB04 = Nesvorn\'y and Bottke (2004), 
T07 = Tsiganis et al. (2007), F06 = Farley et al. (2006), W08 = Willman et al. (2008), 
NV06 = Nesvorn\'y and Vokrouhlick\'y (2006), MH09 = Molnar and Haegert (2009).}
\end{table*}

The age of the Karin family has been established by numerically integrating the orbits of identified 
members back in time in an attempt to identify their past convergence (Fi\-gu\-re \ref{karin}). The past
convergence is expected because the spread of the Karin family in $a_{\rm P},e_{\rm P},i_{\rm P}$
indicates that the ejection speeds of observed fragments were only $\lesssim$15 m s$^{-1}$. These low speeds 
imply a very tight {\it initial} distribution (to within $\simeq$1$^\circ$) of $\lambda$, $\varpi$ and 
$\Omega$. The backward integration showed that the convergence occurred at $5.8\pm0.2$ m.y. ago. 
In addition, the past convergence improved, with $\varpi$ and $\Omega$ of all member orbits converging 
to within a degree, if the backward integration included the Yarkovsky drift (Nesvorn\'y and Bottke, 2004). 
This was used to measure the rate of the Yarkovsky drift for individual members of the family, determine 
their obliquities, and pin down the age of the Karin family to $t_{\rm age}=5.75\pm0.05$ m.y.
    
The method of backward integration of orbits was applied to several families (Nesvorn\'y et al., 2002a, 2003, 2008b; 
Novakovi\'c et al., 2010, 2012a,b, 2014; Table \ref{youngfam}; some of these results will need to be 
verified). One of the interesting results that emerged from these studies is a possible relationship
between the young families and Main Belt Comets (MBCs; see chapter by Jewitt et al. in this volume). For example, (7968) P/133 
Elst-Pizzaro can be linked to the Beagle family ($t_{\rm age}\sim10$ m.y.; Nesvorn\'y et al., 2008b), and (300163) P/2006 
VW139 and P/2012 F5 (Gibbs) can be linked to small families that probably formed within the past 
10 m.y. (Novakovi\'c et al., 2012a, 2014). If this relationship is confirmed by future studies, this can 
help us to understand how the MBCs are ``activated''.   

Another notable case of a recent breakup is the Veritas family (Figure \ref{veritas}). It has previously been 
hypothesized that the Veritas family is $<$50 m.y. old (Milani and Farinella 1994). This claim was 
based on the argument that the largest member of the dynamical family, (490) Veritas, has chaotic dynamics and 
would be expected to diffuse away in $e_{\rm P}$ from the rest of the family, if the family were older 
than $\sim$50~m.y. [Note, however, that recent impact modeling may indicate that (490) Veritas is not a true 
member of the Veritas family (e.g., Michel et al., 2011).] A backward integration of orbits confirmed the young age, 
and showed that the Veritas family formed only $8.3\pm0.5$ m.y. ago (Nesvorn\'y et al., 2003). 

A similarly young age was later obtained by an independent method, known as the ``chaotic chronology'', 
based on tracking the evolution of orbits in one of the diffusive resonances that intersect the Veritas family 
(Tsiganis et al., 2007, see also Kne\v{z}evi\'c and Pavlovi\'c, 2002).
Tsiganis et al. (2007) considered the chaotic diffusion of the Veritas family members in the 
5J-2S-2 three-body resonance at 3.174~AU (Nesvorn\'y and Morbidelli, 1999). Based on numerical integrations 
of chaotic orbits they estimated that the observed spread in the 5J-2S-2 resonance can be obtained for $t_{\rm age}
=8.7\pm1.7$ m.y.  Interestingly, the Veritas family is also intersected by the 7J-7S-2 resonance at 
3.168~AU. The observed distribution of eccentricities in this resonance is rather wide and cannot be
explained by normal diffusion over the estimated age (the 7J-7S-2 resonance is $\sim$100 times less diffusive 
than the 5J-2S-2 resonance). Perhaps the problem is with the HCM chaining, discussed in Section 2.2, 
which links unrelated asteroids in the 7J-7S-2 resonance, or the dynamical modeling is missing some 
important ingredient. Novakovi\'c et al. (2010) applied the method of chaotic chronology to the Theobalda 
family and found that the estimated age is consistent with that obtained from a backward integration 
of the regular orbits ($t_{\rm age}=6.9\pm2.3$~m.y.).          

There is a close relationship of the young asteroid families to the {\it asteroid dust bands} (see chapter
by Jenniskens in this volume), which are 
strips of infrared emission running roughly parallel to the plane of the solar system (Low et al., 
1984). The three most prominent dust bands, known as $\alpha$, $\beta$ and $\gamma$, have previously been  
thought to originate in the Themis, Koronis and Eos families (Dermott et al., 1984). A detailed 
modeling, however, have shown that the sources of these dust bands are the recently-formed Karin, Beagle 
and Veritas families (Dermott et al., 2002; Nesvorn\'y et al., 2003, 2006, 2008b), mainly because: (1) the 
Veritas family with $i_{\rm P}=9.3^\circ$ provides a better fit to the latitudinal position of the 
$\gamma$ band than the Eos family with $i_{\rm P}\simeq10^\circ$, and (2) the young families should now 
be more prolific sources of dust than the old families, because the dust production in a collisional 
cascade is expected to drop with time. A tracer of the Veritas family breakup has been found in 
measurements of extraterrestrial $^3$He in $\simeq$8.2-m.y.-old deep ocean sediments (Farley et al., 2006).       

\begin{figure}[t!]
\epsscale{1.0}
\plotone{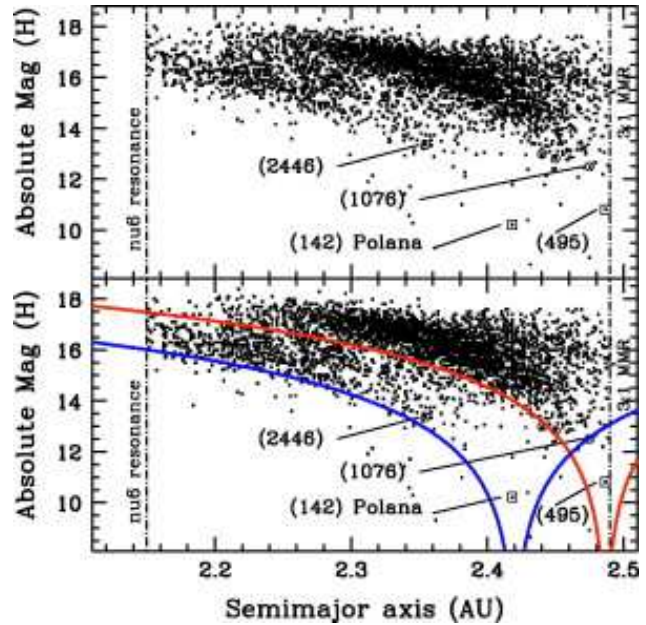}
\caption{\small A recent application of the V-shape criterion to define the Polana and Eulalia families, 
and identify the largest members in these families (Walsh et al., 2013). The plot shows
the absolute magnitude $H$ of dark ($p_V<0.1$) asteroids with $0.1<e_{\rm P}<0.2$ and $i_{\rm P}<10^\circ$ as a 
function of $a_{\rm P}$. The bottom panel illustrates the best-fit V-shape curves to the Polana (left) 
and Eulalia (right) families. (142) Polana and (495) Eulalia, the largest dark asteroids inside the 
left and right zones, respectively, are the likely largest members of the two families.}   
\label{eulalia}
\end{figure}

\section{\textbf{V-SHAPE CRITERION}}
A correct identification of the brightest/largest members in a family is important for several 
reasons. For example, the spectroscopic observations are magnitude limited and can typically 
only be conducted for bright targets. The interpretation of spectroscopic observations of asteroid 
families, and implications for the homogeneity/heterogeneity of their parent bodies, therefore depend 
on whether the bright asteroids are the {\it actual} members of a family, or not (e.g., Reddy et al., 2011). 
The large objects in families are also critically important for asteroid impact studies, in which the 
Size Frequency Distribution (SFD) of family members is used to calibrate the results of impact
experiments. Finally, the uncertainty in the membership of the largest family members critically 
affects our estimates of size of the disrupted parent bodies (e.g., Durda et al., 2007).

While the interloper problem may not have an ideal solution, there exists a straightforward method that 
can be used to remove obvious bright/large interlopers based on dynamical considerations. This method 
is inspired by the ``V-shape'' of asteroid families, which becomes apparent when the absolute magnitude 
$H$ of family members is plotted against $a_P$ (Figure~\ref{eulalia}). The V shape results from  
two processes. First,  larger (smaller) fragments tend to be ejected at lower (higher) speeds, and thus 
tend to be located, on average, closer to (further away from) the family center (see chapter 
by Michel et al. in this volume). The second and typically more dominant 'V' is contributed by the 
Yarkovsky effect. 

The Yarkovsky effect (YE) is a recoil force produced by anisotropic emission of thermal photons from 
an asteroid surface (see chapter by Vokrouhlick\'y et al. in this volume). The diurnal component of the YE, 
which is more important for asteroid-sized bodies than its seasonal counterpart, can increase the semimajor axis 
of an asteroid with a prograde rotation and decrease the semimajor axis for a retrograde rotation. 
The semimajor axis drift is generally given by ${\rm d}a_{\rm P}/{\rm d}t = {\rm const.} \cos \theta/(a_{\rm P}^2 D)$, where 
$\theta$ is the asteroid obliquity, $D$ is the effective diameter, and the constant depends on material properties.
The maximum drift occurs for $\theta=0^\circ$ or $\theta=180^\circ$. Thus, the envelope of the distribution 
of family members in $(a_{\rm P},H)$ is expected to follow $|a_{\rm P}-a_{\rm c}| = C_{\rm YE} 10^{H/5}$, where 
$a_{\rm c}$ is the family center (often assumed to coincide with the largest fragment), and $C_{\rm YE}$ is 
a constant related to $t_{\rm age}$ (Section 5). Interestingly, if the ejection speed $\delta V \propto 1/D$, as 
found for the youngest families (e.g., Nesvorn\'y et al., 2006a) and some laboratory experiments (e.g., Fujiwara 
et al., 1989), then the ejection velocity field will produce the same dependence, $|a_{\rm P}-a_{\rm c}| = 
C_{\rm EV} 10^{H/5}$, where $C_{\rm EV}$ is related to the magnitude of the ejection velocities. 
 
These considerations allow us to define the V-shape criterion (e.g., Nesvorn\'y et al., 2003). Consider a 
family extracted by the HCM as described in Section 2.2. Fit an envelope to the distribution of small family 
members in $(a_{\rm P},H)$ using the functional dependence between $a_{\rm P}$ and $H$ defined above. The 
envelope is then continued to low $H$ values, and the bright members of the HCM family that fall {\it outside} the 
envelope boundaries can be rejected. On the other hand, the brightest HCM family members that fall 
{\it within} the envelope boundaries are good candidates for the largest family members. This method 
is~illustrated~in~Figure~\ref{eulalia}. 

In practice, a number of additional effects can complicate the application of the V-shape criterion 
described above. For example, strong nearby resonances can remove family members that drifted into 
them, thus producing cutoffs of the $a_{\rm P}-a_{\rm c}$ distribution beyond which no family members can be 
identified (e.g., the 3:1 resonance in Figure \ref{eulalia}). Also, family members can be displaced in 
$a_{\rm P}$ by encounters with (1) Ceres and other massive main belt asteroids (Nesvorn\'y et al., 2002b; 
Carruba et al., 2003, 2007a, 2012, 2013b; Delisle and Laskar, 2012). In addition, the physics of large scale collisions is still poorly 
understood, and the possibility that some large fragments can be accelerated to very high speeds cannot 
be ruled out. Therefore, while the V-shape criterion defined above is a useful guide, it cannot be 
{\it rigidly} applied. 

Possibly the best way to deal with this issue is to define the value of $C_0 = 10^{-H/5} |a_{\rm P}-a_{\rm c}|$ 
that best fits the V-shaped family envelope and report $C_j/C_0$, where $C_j= 10^{-H_j/5} |a_{{\rm P},j}-a_{\rm c}|$,
for each family member $j$ identified by the HCM. Asteroids with $|C_j/C_0|>1$ can 
be flagged (but not removed), because they are potential 
interlopers in a dynamical family according to the V-shape criterion. The results can then be cross-linked 
with the spectroscopic data (or colors, or albedo measurements) to determine whether there is a good correspondence 
between the flagged bodies and (suspected) spectroscopic interlopers. An illustration of this procedure can 
be found in Vokrouhlick\'y et al. (2006a) who examined the Eos family. They found that many large bodies in 
the HCM family with $|C_j/C_0|>1$ have physical properties that are indeed incompatible (mainly dark C types) 
with the bulk of the Eos family (mainly brighter K types).  
\section{\textbf{FAMILY AGE ESTIMATION}}
The method of backward integrations of orbits described in Section 3 cannot be used to determine 
ages of the families much older than $\sim$10 m.y., mainly because the orbital evolution of main belt asteroids is 
generally unpredictable on long timescales and sensitively depends on non-gravitational effects 
that are difficult to model with the needed precision. Instead, the age of an old family can be 
estimated by a {\it statistical} method, which is based on the general notion that the spread of an 
asteroid family in $a_{\rm P}$ increases over time as its members drift away due to the Yarkovsky effect. 
Expressed in terms of the equations discussed in the previous section, the age can be estimated~as
\begin{eqnarray} 
t_{\rm age} \simeq 1\, {\rm g.y.} 
\!\!\!& \times & \!\!\!\left({C_0 \over 10^{-4}{\rm AU}}\right)
\left({a \over 2.5\,{\rm AU}}\right)^2 \nonumber \\  
\!\!\!&\times &\!\!\!  \left({\rho \over 2.5\,{\rm g\, cm^{-3}}}\right)  
\left({0.2 \over p_{\rm V}}\right)^{1 \over 2} \,
\label{tage} 
\end{eqnarray}
where $\rho$ is the asteroid bulk density and $p_{\rm V}$ is the visual albedo. While the equation above is 
scaled to typical values expected for an S-type asteroid, a change to $\rho=1.5$ g cm$^{-3}$ and 
$p_{\rm V}=0.05$, which would be more appropriate for a dark C-type asteroid, produces two multiplication 
factors that nearly compensate each other. The biggest uncertainty in the inversion from $C_0$ to $t_{\rm age}$ 
lies in the unknown density factor. Additional uncertainty, not explicitly apparent from Eq. (2), stems from 
the dependence of the drift rate on surface conductivity $K$. Together, the conversion from $C_0$ to $t_{\rm age}$ 
has an uncertainty of about a factor of 2. 

More fundamentally, Eq. (\ref{tage}) neglects complicating factors such as the contribution of the 
original ejection field and nearby resonances that can remove drifting bodies. If $C=C_{\rm YE}+C_{\rm EV}$,
under the assumptions on the ejection speeds discussed in Section 4, the two effects cannot be decoupled from each
other, and $t_{\rm age}$ estimated from $C_0$ will always overshoot the real age of the family. This happens,
in essence, because this simple method only fits the family envelope.

Vokrouhlick\'y et al. (2006a,b) have developed a more general statistical method that uses the actual 
distribution of family members within the V-shape envelope. It works as follows. First, the code generates a 
new-born family. The distribution of the ejection speeds of fragments is approximated by the Gaussian distribution 
in each velocity component with the size-dependent standard deviation $\delta V=V_5(5\ {\rm km}/D)$, where 
$V_5$ is a free parameter. This functional dependence generally provides a good match to the ejection speeds 
inferred from the young  families and impact simulations (Michel et al., 2001, 2003, 2004; Nesvorn\'y et al., 
2002a, 2006a; Durda et al., 2004, 2007). The Yarkovsky-YORP (hereafter YY) code then evolves the orbits of 
fragments accounting for the semimajor axis drift due to the YE, and the spin evolution from the YORP effect 
(Rubincam, 2000). To speed up the calculation, planetary perturbations were not taken into account in the 
original YY code (but see Masiero et al., 2012; results based on full $N$-body integrations are discussed
in the following section). 

\begin{figure*}
\epsscale{1.5}
\plotone{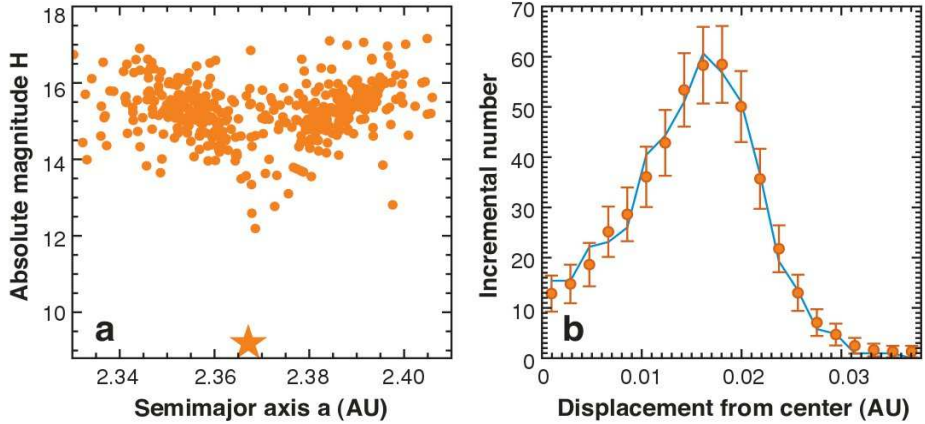}
\caption{\small 
(a) The Erigone family projected onto a plane of proper semimajor axis versus absolute
magnitude $H$; (163) Erigone is the filled star. The family has been separated into two clouds 
($a_{\rm p}\lesssim 2.37$ AU; $a_{\rm p}\gtrsim 2.37$ AU) by the Yarkovsky/YORP evolution. (b) A comparison 
between model results (solid line) and binned Erigone family (gray dots; see Vokrouhlick\'y 
et al., 2006b). The error bars are the square root of the number of bodies in each bin. The $x$-axis is 
the distance of family members from the the family center. Based on this result,
Vokrouhlick\'y et al. (2006b) estimated that $t_{\rm age}=280^{+30}_{-50}$ m.y. and $V_5=26^{+14}_{-11}$ 
m s$^{-1}$, where the error bars do {\it not} include the uncertainty originating from uncertain material 
properties (e.g., density, surface conductivity).} 
\label{erigone}
\end{figure*}

When applied to an asteroid family, the goal is to find a combination of parameters $V_5$ and 
$t_{\rm age}$ that best fits the observed $(a_{\rm P},H)$ distribution in the family. Since the 
model does not contain an SFD-related evolution component, this can be conveniently done by simply 
fitting the observed distribution of $C_j$. The results are found to be credible in cases such as the 
Erigone family (Figure \ref{erigone}), where the YY model is capable of adequately representing the
observed $C$ distribution, which shows void space near $C=0$, a maximum at intermediate values of 
$C$, and a relatively sharp drop toward the largest $C$ values seen in the family. This dependence
is produced by the YORP effect which tilts the spin axis away from the orbital plane and therefore,
through the $\cos \theta$ dependence of the YE, tends to maximize the drift rates.    
This gives some asteroid families the characteristic appearance in $(a_{\rm P},H)$, which different 
authors, with noticeably different gifts for subtleties of poetic expression, called ``ears'', ``wings'' 
or ``petals''.   
    
But not all families have ears, wings or petals. In fact, most families have more uniform 
distribution of $C$ that goes all the way from 0 to $C_{\rm max}$. This is thought to have something 
to do with how the YE and YORP operate on g.y.-long timescales. It is perhaps related 
to the variability of the spin-related YORP torque, which sensitively depends on small perturbations 
of the asteroid surface. As a result of this dependence, the spin rate of each individual asteroid can 
undergo a random walk (stochastic YORP; Bottke et al., 2015) and, when spun down completely, the asteroid 
spin axis would chaotically tumble for some time, and subsequently reorient. This process could mix up the 
total drifts suffered by family members and, rather unpoetically, remove the ears, wings or petals from 
the old families. 

The application of the standard YY code (and, for that matter, also of the simple method based on the 
`V'-shape envelope) is problematic in cases when a family is cut by nearby mean motion resonances and 
looks like a box in $(a_{\rm P},H)$, or when $C_{\rm EV}$ is expected to be larger than $C_{\rm YE}$,
perhaps because the family's parent was an object with large escape speed (e.g., the Vesta family; 
(4) Vesta has $V_{\rm esc} \simeq 360$ m s$^{-1}$), or because the family formed recently
(e.g., the Datura, Karin, Veritas families have $C_{\rm YE}\simeq0$). A complete list of families to 
which the YY code was successfully applied so far is:
\begin{itemize} 
\item Agnia ($t_{\rm age}=100\pm100$ m.y., $V_5\simeq15$ m s$^{-1}$)
\item Massalia ($t_{\rm age}=150\pm50$ m.y., $V_5\simeq20$ m s$^{-1}$)
\item Baptistina ($t_{\rm age}=160\pm50$ m.y., $V_5\simeq40$ m s$^{-1}$)
\item Merxia ($t_{\rm age}=250\pm100$ m.y., $V_5\simeq25$ m s$^{-1}$) 
\item Astrid ($t_{\rm age}=250\pm100$ m.y., $V_5\simeq15$ m s$^{-1}$) 
\item Erigone ($t_{\rm age}=300\pm100$ m.y., $V_5\simeq30$ m s$^{-1}$)
\item Eos ($t_{\rm age}=1.3\pm0.5$ g.y., $V_5\simeq70$ m s$^{-1}$)\ 
\item Tina ($t_{\rm age}=170\pm50$ g.y., $V_5\simeq20$ m s$^{-1}$)\ . 
\end{itemize}

Here we have taken the liberty to update and round off the estimates, and give more generous errors
than in the original publications (Vokrouhlick\'y et al., 2006a,b,c; Bottke et al., 2007; Carruba and Morbidelli, 
2011; see also Carruba, 2009a for the Padua family). Note that these errors do not include the uncertainty of about 
a factor of 2 from the poorly known bulk density and surface conductivity of the asteroids in question. 
Including this uncertainty, Masiero et al. (2012) found that the best-fitting age of the Baptistina family 
can be anywhere between 140 and 320 m.y.

The estimated ejection speeds are $V_5=15$-50 m s$^{-1}$, except for the Eos family, which
formed in a breakup of a very large parent asteroid ($D_{\rm PB}\sim300$~km). These results
are consistent with the ejection speeds inferred from the young Karin family, which has 
$V_5\simeq15$~m~s$^{-1}$ for a relatively small parent body ($D_{\rm PB}\simeq 35$ km; Nesvorn\'y 
et al., 2006a). The ejection speeds contribute by $\simeq$20\% 
(for oldest Eos) to 50\% (for youngest Agnia) to the total family spread in the semimajor axis.
Ignoring this contribution, as in Eq. (2), would thus lead to an overestimate of $t_{\rm age}$ 
by $\simeq$20-50\%. While one must therefore be careful in applying Eq. (2) to the small/young 
families that did not have enough time to significantly spread by the YE, the effect of the ejection 
speeds should be less of an issue for old families. 
\section{\textbf{DYNAMICAL EVOLUTION}}
\smallskip
\noindent
\textbf{6.1 INITIAL STATE}
\bigskip

The dynamical evolution of asteroids in families is similar to the dynamical evolution of main belt 
asteroids in general. Studying the dynamical evolution of individual families is useful in this 
context, because we more or less know how the families should look like initially. Things may thus be 
learned by comparing these ideal initial states with how different families look now, after having
dynamically evolved since their formation. The dynamical studies can also often provide an independent 
estimate of $t_{\rm age}$. 
 
Assuming that $\delta V \ll V_{\rm orb}$, the initial shape of families in $(a,e,i)$ can be obtained from 
the Gauss equations (e.g., Zappal\`a et al., 2002), which map the initial velocity perturbation 
$\mathbf{\delta V}=(V_{\rm R},V_{\rm T},V_{\rm Z}$), where $V_{\rm R}$, $V_{\rm T}$ and $V_{\rm Z}$ are the
radial, tangential and vertical components of the velocity vector, to the change of orbital elements 
$\mathbf{\delta E}=(\delta a,\delta e,\delta i)$. If the ejection velocity field is (roughly) isotropic,
the Gauss equations imply that initial families should (roughly) be ellipsoids in $(a,e,i)$ centered 
at the reference orbit $(a^*,e^*,i^*)$. The transformation from $(a,e,i)$ to $(a_{\rm P},e_{\rm P},i_{\rm P})$ 
preserves the shape, but maps $(a^*,e^*,i^*)$ onto $(a_{\rm P}^*,e_{\rm P}^*,i_{\rm P}^*)$ such that, 
in general, $a_{\rm P}^* \neq a^*$, $e_{\rm P}^* \neq e^*$ and $i_{\rm P}^* \neq i^*$.
 
The shape of the ellipsoids in $(a_{\rm P},e_{\rm P},i_{\rm P})$  is controlled by the true anomaly $f$ and 
the argument of perihelion $\omega$ of the parent body at the time of the family-forming breakup. The 
projected distribution onto the $(a_{\rm P},e_{\rm P})$ plane is a tilted ellipse with tightly correlated 
$a_{\rm P}$ and $e_{\rm P}$ if the breakup happened near perihelion (see Figure \ref{karin}a) or tightly 
anticorrelated $a_{\rm P}$ and $e_{\rm P}$ if the breakup happened near aphelion. The two recently-formed families 
for which this shape is clearly discernible, the Karin and Veritas families, have correlated $a_{\rm P}$ 
and $e_{\rm P}$, implying that $|f| \simeq 30^\circ$ (Figures \ref{karin} and \ref{veritas}). 

The projected initial distribution onto the $(a_{\rm P},i_{\rm P})$ plane
is an ellipse with horizontal long axis and vertical short axis. The short-to-long axis ratio is 
roughly given by $\cos(\omega+f)V_{\rm Z}/V_{\rm T}$. Thus, breakups near the ascending ($\omega+f\simeq 0$) 
and descending ($\omega+f\simeq \pi$) nodes should produce `fat' ellipses while those with 
$\omega+f=\pm\pi/2$ should make `squashed' ellipses with $\delta i_{\rm P} \simeq 0$. While the Karin
family neatly fits in this framework with $\omega+f \simeq \pi/4$ (Fi\-gu\-re \ref{karin}), the Veritas 
family shows large $\delta i_{\rm P}$ values, indicating that the ejection velocity field should
have been anisotropic with $V_{\rm Z}$ some $\simeq2$-4 times larger than $V_{\rm T}$.

\begin{figure*}[t]
\centering
\includegraphics[width=17cm]{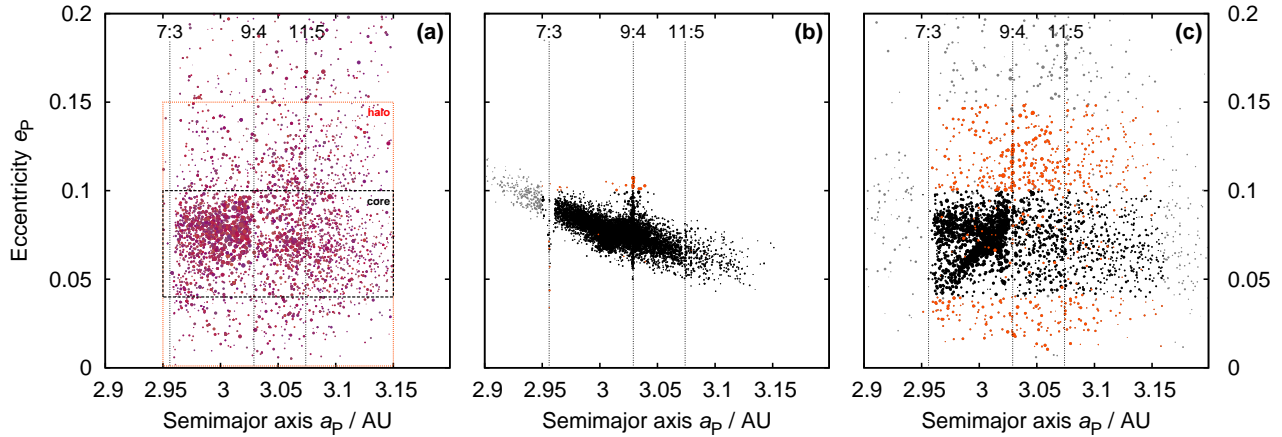}
\caption{\small Dynamical evolution of the Eos family. From left to right, the panels show the: (a) observed
family and its halo in the $(a_{\rm P},e_{\rm P})$ projection, (b) assumed initial shape of the family, 
and (c) family's structure after 1.7 g.y. In (a), we plot all asteroids with Eos-family colors 
($0.0<a^*<0.1\,{\rm mag}$ and $-0.03<i-z<0.08\,{\rm mag}$; see Ivezi\'c et al. (2001) for the definition
of color indices from the SDSS). The size of a symbol is inversely proportional to absolute magnitude $H$. 
The boxes approximately delimit the extent of the core and halo of the Eos family.
In (b), 6545 test particles were distributed with assumed isotropic ejection velocities, 
$V_5 = 93$~m~s$^{-1}$, $f = 150^\circ$ and $\omega = 30^\circ$. Nearly all initial particles fall within the
family core. In (c), an $N$-body integrator was used to dynamically evolve the orbits of the test particles 
over 1.7~g.y. The integration included gravitational perturbations from planets, and the Yarkovsky
and YORP effects. The vertical lines show the locations of several resonances that contributed to spreading of orbits
in $e_{\rm P}$ (7:3, 9:4 and 11:5 with Jupiter, also $3{\rm J}$-$2{\rm S}$-$1$ and $z_1 \equiv g+s-g_6-s_6=0$). 
Adapted from Bro\v{z} and Morbidelli (2013).}
\label{eos}
\end{figure*}

The reference orbit $(a_{\rm P}^*,e_{\rm P}^*,i_{\rm P}^*)$ is often taken to coincide with the proper orbit 
of a largest family member. This should be fine for families produced in cratering or mildly catastrophic 
events, where the orbital elements of the impacted body presumably did not change much by the impact.  
For the catastrophic and highly catastrophic breakups, however, the largest surviving remnant is relatively 
small and can be significantly displaced from the family's center.
For example, (832) Karin, the largest $\simeq$17-km-diameter member of the Karin family produced by a 
catastrophic breakup of a $\simeq$40-km-diameter parent body (mass ratio $\sim$0.08; Nesvorn\'y et al.,
2006a), is displaced by $-0.002$~AU from the family center ($\simeq$20\% of the whole extension of the 
Karin family in $a_{\rm P}$). This shows that, in general, the position of the largest fragment does not 
need to perfectly coincide with the family center, and has implication for the V-shape criterion discussed 
in Section 4 (where an allowance needs to be given for a possible displacement).   

\bigskip
\noindent
\textbf{6.2 DYNAMICS ON GIGAYEAR TIMESCALES}
\bigskip

An overwhelming majority of the observed asteroid families are not simple Gaussian ellipsoids. While this was 
not fully appreciated at the time of the Asteroids III book, today's perspective on this issue is clear: 
{\it the families 
were stretched in $a_{\rm P}$ as their members drifted away from their original orbits by the Yarkovsky 
effect}. The asteroid families found in the present main belt are therefore nearly horizontal and 
elongated structures in $(a_{\rm P},e_{\rm P})$ and $(a_{\rm P},i_{\rm P})$. This shows that 
the original ejection velocity field {\it cannot} be easily reconstructed by simply mapping back today's 
$(a_{\rm P},e_{\rm P},i_{\rm P})$ to $(V_{\rm R},V_{\rm T},V_{\rm Z})$ from the Gauss equations (Note 7).

Moreover, many asteroid families have weird shapes which, taken at the face value, would imply funny and
clearly implausible ejection velocity fields. A prime example of this, briefly mentioned in section 2.2, is the 
Koronis family (Bottke et al., 2001). Since the case of the Koronis family was covered in the Asteroids 
III book (chapter by Bottke et al., 2002), we do not discuss it here. Instead, we concentrate on 
the results of new dynamical studies, many of which have been inspired by the Koronis family case. 
The dynamical effects found in these studies fall into three broad categories: 

(1) Members drifting in $a_{\rm P}$ encounter a mean motion resonance with one of the planets (mainly Jupiter, 
Mars or Earth; see chapter Nesvorn\'y et al., 2002c in Asteroids III book). If the resonance is strong enough (e.g., 
3:1, 2:1, or 5:2 with Jupiter), the orbit will chaotically wander near the resonance border, its eccentricity
will subsequently increase, and the body will be removed from the main belt and transferred onto a planet-crossing 
orbit (Wisdom, 1982). If the resonance is weak, or if the asteroid is small and drifts fast in $a_{\rm P}$, the orbit 
can cross the resonance, perhaps suffering a discontinuity in $e_{\rm P}$ during the crossing, and will continue drifting on 
the other side. If the resonance is weak and the drift rate is not too large, the orbit can be captured in the resonance 
and will slowly diffuse to larger or smaller eccentricities. It may later be released from the resonance 
with $e_{\rm P}$ that can be substantially different from the original value. The effects of mean 
motion resonances on $i_{\rm P}$ are generally smaller, because the eccentricity terms tend to be more important 
in the resonant potential. The inclination terms are important for orbits with $i_{\rm P}\gtrsim 10^\circ$. 
A good example of this is the Pallas family with $i_{\rm P}\simeq33^\circ$ (Carruba et al., 2011).

(2) Drifting members meet a secular resonance. The secular resonances are located along curved
manifolds in $(a_{\rm P},e_{\rm P},i_{\rm P})$ space (Kne\v{z}evi\'c et al., 1991). Depending on the 
type and local curvature of the secular resonance, and asteroid's ${\rm d}a/{\rm d}t$, the orbit can be 
trapped inside the resonance and start sliding along it, or it can cross the resonance with a noticeably
large change of $e_{\rm P}$ and/or $i_{\rm P}$. A good example of the former case are orbits in the 
Eos family sliding along the $z_1=g-g_6+s-s_6=0$ resonance (Vokrouhlick\'y et al., 2006a). An example of 
the latter case is the Koronis family, where eccentricities change as a result of crossing of 
the $g+2g_5-3g_6=0$ resonance (Bottke et al., 2001). If the secular resonance in question only includes the
$g$ (or $s$) frequency, effects on $e_{\rm P}$ (or $i_{\rm P}$) are expected. If the resonance includes both 
the $g$ and $s$ frequencies, both $e_{\rm P}$ and $i_{\rm P}$ can be effected. If the orbit is captured in a 
resonance with the $g$ and $s$ frequencies, it will slide along the local gradient of the resonant 
manifold with changes of $e_{\rm P}$ and $i_{\rm P}$ depending on the local geometry.

(3) Encounters with (1) Ceres and other massive asteroids produce additional changes of $a_{\rm P}$, $e_{\rm P}$ 
and $i_{\rm P}$ (Nesvorn\'y et al., 2002b; Carruba et al., 2003, 2007a, 2012, 2013b; Delisle and Laskar, 2012). These changes 
are typically smaller than those from the Yarkovsky effect on $a_{\rm P}$ and resonances on $e_{\rm P}$ and $i_{\rm P}$. 
They are, however, not negligible. The effect of encounters can be approximated by a random walk (see Carruba et al., 
2007a for a discussion).
The mean changes of $a_{\rm P}$, $e_{\rm P}$ and $i_{\rm P}$ increase with time roughly as $\sqrt{t}$. The asteroid families become 
puffed out as a result of encounters, and faster so initially than at later times, because of the nature of 
the random walk. Also, a small fraction of family members, in some cases perhaps including the largest 
remnant, can have their orbits substantially affected by a rare, very close encounter. Additional perturbations 
of asteroid orbits arise from the linear momentum transfer during non-disruptive collisions (Dell'Oro and 
Cellino, 2007). 
     
\bigskip
\noindent
\textbf{6.3 DISCUSSION OF DYNAMICAL STUDIES}
\bigskip

The Koronis family case (Bottke et al., 2001) sparked much interest in studies of the dynamical evolution of asteroid
families on very long timescales. Here we review several of these studies roughly in the chronological order. 
The goal of this text is to illustrate the dynamical processes discussed in the previous section on specific 
cases.

Nesvorn\'y et al. (2002b) considered the dynamical evolution of the Flora family. The Flora family is located 
near the inner border of the main belt, where numerous mean motion resonances with Mars and Earth produce 
slow diffusion of $e_{\rm p}$ and $i_{\rm p}$. The numerical integration of orbits showed how the overall 
extent of the Flora family in $e_{\rm P}$ and $i_{\rm P}$ increases with time. The present width of the Flora family 
in $e_{\rm P}$ and $i_{\rm P}$ was obtained in this study after $t \simeq 0.5$ g.y. even if the initial 
distribution of fragments in $e_{\rm P}$ and $i_{\rm P}$ was very tight. The Flora family expansion saturates 
for $t>0.5$ g.y., because the Flora family members that diffused to large eccentricities are removed from
the main belt by encounters with Mars (the Flora family is an important source of chondritic near-Earth 
asteroids (NEAs); Vernazza et al., 2008). The present spread of the Flora family in $a_{\rm P}$, mainly 
contributed by the Yarkovsky effect, indicates $t_{\rm age}\sim1$ g.y. 
 
Carruba et al. (2005) studied the dynamical evolution of the Vesta family. The main motivation for this study
was the fact that several inner main belt asteroids, such as (956) Elisa and (809) Lundia, have been classified 
as V-types from previous spectroscopic observations (Florczak et al., 2002), indicating that they may be pieces 
of the basaltic crust of (4) Vesta. These asteroids, however, have orbits rather distant form that of (4) Vesta 
and are not members of the Vesta's dynamical family even if a very large cutoff distance is used. It was 
therefore presumed that they: (i) have dynamically evolved to their current orbits from the Vesta family,  
or (ii) are pieces of differentiated asteroids unrelated to (4) Vesta. Carruba et al. (2005) found that
the interplay of the Yarkovsky drift and the $z_2\equiv 2(g-g_6)+s-s_6=0$ resonance 
produces complex dynamical behavior that can, indeed, explain the orbits of (956) Elisa and (809) Lundia, 
assuming that the Vesta family is at least $\simeq$1 g.y. old. This gives support to (i). 

In a follow-up study, Nesvorn\'y et al. (2008a) performed a numerical integration of 6,600 Vesta fragments over 
2 g.y. They found that most V-type 
asteroids in the inner main belt can be explained by being ejected from (4)~Vesta and dynamically evolving 
to their current orbits outside the Vesta family. These V-type ``fugitives'' have been used to constrain the 
age of the Vesta family, consistently with findings of Carruba et al. (2005), to $t_{\rm age}\gtrsim1$~g.y. 
Since previous collisional modeling of the Vesta family suggested $t_{\rm age}\lesssim 1$ g.y. (Marzari et al., 
1999), the most likely age of the Vesta family that can be inferred from these studies is $t_{\rm age} 
\sim 1$~g.y. This agrees well with the age of the $\simeq$500-km-diameter Rheasilvia basin on (4) Vesta inferred 
from crater counts ($\simeq1$ g.y.; Marchi et al., 2012) (Note 8).

Vokrouhlick\'y et al. (2006a) studied the dynamical evolution of the Eos family. The Eos family has a 
complicated structure in proper element space leading some authors to divide it in several distinct families 
(e.g., Milani et al., 2014). Diagnostically, however, the Eos family, although somewhat physically 
heterogeneous, has the color, albedo and spectral properties that contrast with the local, predominantly 
C-type background in the outer asteroid belt. This suggests that this is a single family. As we 
discuss below, the complicated structure of the Eos family arises from of the presence of several mean 
motion and secular resonances. 

To start with, Vokrouhlick\'y et al. (2006a) showed that the Eos family members drifting by the 
Yarkovsky effect into the 7:3 resonance with Jupiter are removed (see Figure~\ref{eos}). This cuts 
the family at 2.957 AU. 
Members drifting with ${\rm d}a/{\rm d}t>0$, on the other hand, will encounter the 9:4 resonance at
3.03 AU. This resonance, being of higher order and thus weaker, is not an unpenetrable barrier, 
especially for smaller members with higher drift rates. The estimated fraction of bodies that can 
cross the 9:4 resonance is $<$10\% for $H<12$ but reaches $\simeq$35\% for $H=16$. This is consistent 
with the magnitude distributions of the Eos family members on both sides of the 9:4 resonance. 
The larger dispersion of the part of the Eos family with $a_{\rm P}>3.03$ AU is contributed by 
perturbations of $e_{\rm P}$ and $i_{\rm P}$ during the 9:4 resonance crossing (Figure \ref{eos}). Finally, many 
orbits in the central part of the Eos family are trapped in the secular resonance $z_1\equiv g+s-g_6-s_6=0$, 
and slide along it while drifting in $a_{\rm P}$ (Note 9).

Finally, we discuss additional processes whose significance is shadowed by the Yarkovsky effect 
and resonances, but which can be important in some cases.
Nesvorn\'y et al. (2002b) considered encounters with (1) Ceres and found that the characteristic 
change of the semimajor axis due to these encounters is $\Delta a \simeq 0.001$~AU over 100 m.y. Assuming that the 
scattering effect of encounters can be described by a random walk with $\Delta a \propto \sqrt{t}$, the expected 
changes over 1 g.y. and 4 g.y. are $\simeq$0.003 AU and $\simeq$0.007 AU, respectively. The orbital changes were 
found to be larger for orbits similar to that of (1) Ceres, because the orbital proximity leads to lower 
encounter speeds, and larger gravitational perturbations during the low-speed encounters. Carruba et al. (2003) 
studied the effect of encounters on the Adeona and Gefion families, both located near (1) Ceres in proper 
element space. They found that the semimajor axis of members of the Adeona and Gefion families can change 
by up to $\sim$0.01 AU over the estimated age of these families.  
 
With similar motivation, Carruba et al. (2007a) considered the effect of encounters of the Vesta family members 
with (4) Vesta. They found the characteristic changes $\Delta a=0.002$ AU, $\Delta e = 0.002$ and 
$\Delta i=0.06^\circ$ over 100~m.y. In a follow-up work, Delisle and Laskar (2012) included the effects of eleven 
largest asteroids. They showed that encounters of the Vesta family members with (4) Vesta and (1) Ceres are 
dominant, contributing roughly by 64\% and 36\% to the total changes, respectively. The functional dependence 
$\Delta a=1.6 \times 10^{-4}\sqrt{t/1\, {\rm m.y.}}\,{\rm AU}$ was used in this work to extrapolate the 
results to longer time intervals. Moreover, Carruba et al. (2013b) studied the influence of these effects 
on the Pallas, Hygiea and Euphrosyne families. They showed that the effects of (2) Pallas --the 3rd most 
massive main-belt asteroid-- on the Pallas family are very small, because these asteroids have high orbital 
inclinations ($i_{\rm P}\simeq33^\circ$), lower frequency of encounters and higher-than-average encounter speeds.   

Dell'Oro and Cellino (2007) pointed out that orbits of main belt asteroids can change as a result
of the linear momentum transfer during non-destructive collisions. They found that the expected semimajor axis 
change from these collisions 
for a $D=50$-km main-belt asteroid is $\Delta a \sim 10^{-4}$~AU over 100~m.y. (with the scaling laws from 
Benz and Asphaug, 1999). This is an order of magnitude lower than the change expected from close encounters with 
large asteroids and comparable to the sluggish drift rate expected from the Yarkovsky effect for $D=50$ km. For $D<50$~km, 
the orbital changes from non-destructive collisions sensitively depend on several unknown parameters, 
such as the SFD of sub-km main-belt asteroids, but the general trend is such that $\Delta a$ drops with 
decreasing $D$ (assuming that the cumulative SFD index is $<$4; Dell'Oro and Cellino, 2007). Since
$\Delta a$ is independent of $D$ for encounters with (1) Ceres, and $\Delta a \propto 1/D$ for the Yarkovsky 
force, these two effects outrun the non-destructive collisions for $D<50$~km. This limits the significance 
of non-destructive collisions for the dynamical evolution of asteroid families. Their effect on $e_{\rm P}$ and $i_{\rm P}$ 
is also minor.

\bigskip
\noindent
\textbf{6.4 FAMILY HALOS}
\bigskip

\begin{figure}[t!]
\epsscale{1.0}
\plotone{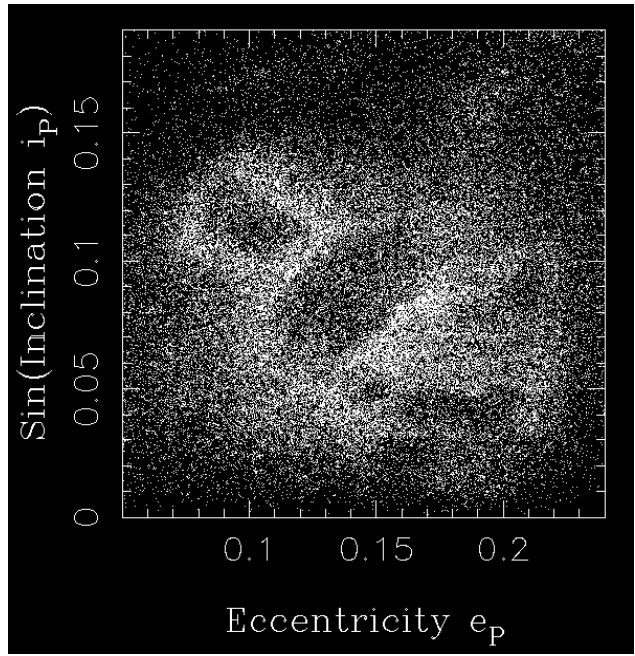}
\caption{\small The halos of families in the inner main belt ($2<a_{\rm P}<2.5$ AU). The dynamical families 
were identified by the HCM in 3D of 
proper elements and were removed, leaving behind what resembles a fuzzy chalk contour of victim's body 
removed from a crime scene. The head, body and bottom leg correspond to the locations of the Vesta, Flora and 
Nysa-Polana families. The halos surrounding these families, mainly composed of small asteroids, are 
rather dramatic in this projection. This figure illustrates a major weakness of the traditional family-identification 
method based on the HCM in 3D space of proper elements.}   
\label{halos}
\end{figure}

When dynamical families are identified and removed from the main belt, they leave behind holes 
in the distribution of proper elements that are surrounded by regions with increased asteroid density 
(Figure \ref{halos}). These peripheral regions are known as the family {\it halos}. The families and family halos 
surrounding them are clearly related, which can most conveniently be demonstrated by considering 
their physical properties. For example, the Koronis family and its halo consist of
bright asteroids (mean albedo $p_{\rm V}=0.15$) that are classified as S in the asteroid taxonomy 
(moderate spectral slope and shallow absorption band near 1 $\mu$m). These properties contrast with 
the local background in the outer main belt, which is mostly dark ($p_{\rm V}\simeq 0.05$) and C-type 
(featureless neutral spectrum). The Eos family and its halo (K-type, $p_{\rm V}=0.13$) also stand out from the
dark outer belt background.   

The distinction between families and family halos is, in a sense, formal, because it 
appears as a consequence of the identification method. Indeed, to extract the family halo by applying
the HCM in 3D space of proper elements, one would need to increase the value of the $d_{\rm cut}$ parameter. 
If it is increased, however, the identification would fail due to the problem of chaining (Section 2.2).
This is a consequence of the fact that the number density of halo asteroids is comparable to, 
or only slightly larger than, the number density of background asteroids. A better method for extracting 
the families jointly with their halos consists in applying the HCM in space of extended dimensions, where 
physical properties are used in addition to the proper elements (Section 2.5; Parker et al., 2008; Carruba et al., 
2013a; Masiero et al., 2013).

The origin of family halos is contributed by at least two processes. In the case of the Eos 
family, the observed dispersion of halo in $e_{\rm P}$ and $i_{\rm P}$ is too large (0-0.15 and 
7-14$^\circ$, respectively) to be related to the the ejection speeds. Instead, Bro\v{z} 
and Morbidelli (2013) showed how the Eos family halo gradually appears as a result of complex orbital 
dynamics in the Eos family region (Figure \ref{eos}). The age of the Eos family inferred
from that work, $1.5\ {\rm g.y.} <t_{\rm age} <2.2$~g.y., is larger than the one obtained from the YY 
code ($t_{\rm age}=1.3 \pm 0.2$ g.y.; Vokrouhlick\'y et al., 2006a and Section~5), but the two estimates 
are consistent at the 1$\sigma$ level.

The Koronis family and its halo, on the other hand, are located in a dynamically quiet region of the main belt 
($2.82<a_{\rm P}<2.95$ AU, $e_{\rm P}\simeq0.045$, $i_{\rm P}\simeq2^\circ$) where no major resonances exist
(except for $g+2g_5-3g_6=0$; Bottke et al., 2001). The long-term dynamics of the Koronis family therefore
has a relatively small influence on the overall width of the family in $e_{\rm P}$ and $i_{\rm P}$. In contrast, 
the present shift of the Koronis family/halo members from the family's center is relatively large
(up to $\simeq0.015$ in $e_{\rm P}$ and $0.5^\circ$ in $i_{\rm P}$). This 
can plausibly be explained by the ejection speeds. 

Assuming an isotropic ejection velocity field and $\delta V = V_5(5\ {\rm km}/D)$ with $V_5=50$ m s$^{-1}$, 
we find that $\delta a_{\rm P}/a_{\rm P} \simeq 2\delta V/V_{\rm orb} \simeq 0.018$ for $D=5$ km fragments. 
This represents a small share of the Koronis family width in $a_{\rm P}$ (the population of $D=5$ km members 
stretches from the 5:2 resonance at 2.82 AU to the 7:3 resonance at 2.95 AU), consistently with the idea 
that the observed spread in $a_{\rm P}$ is mainly due to the Yarkovsky effect. 
With $D=1.5$ km, which is roughly the size of the smallest asteroids observed in this part of the main belt, 
the expected changes are $\delta e_{\rm P} = (1$-$2)\times \delta V/V_{\rm orb} \simeq 0.015$ and $\delta i_{\rm P} 
\lesssim \delta V/V_{\rm orb} \simeq 0.6^\circ$. This is comparable to the present width of the Koronis 
family/halo in $e_{\rm P}$ and $i_{\rm P}$ 

Another notable argument in this context is that the Koronis halo is populated by small asteroids with 
diameters near the minimum detectable size (while the big ones are inside the family core). This suggest a 
certain size dependency of the process that created the halo, presumably related to the size dependency 
of the ejection speeds (see chapter by Michel et al. in this volume). This important trend was pointed out 
by Cellino et al. (2004), and was further discussed in Cellino et al. (2009). 

\begin{figure*}[t!]
\epsscale{1.7}
\plotone{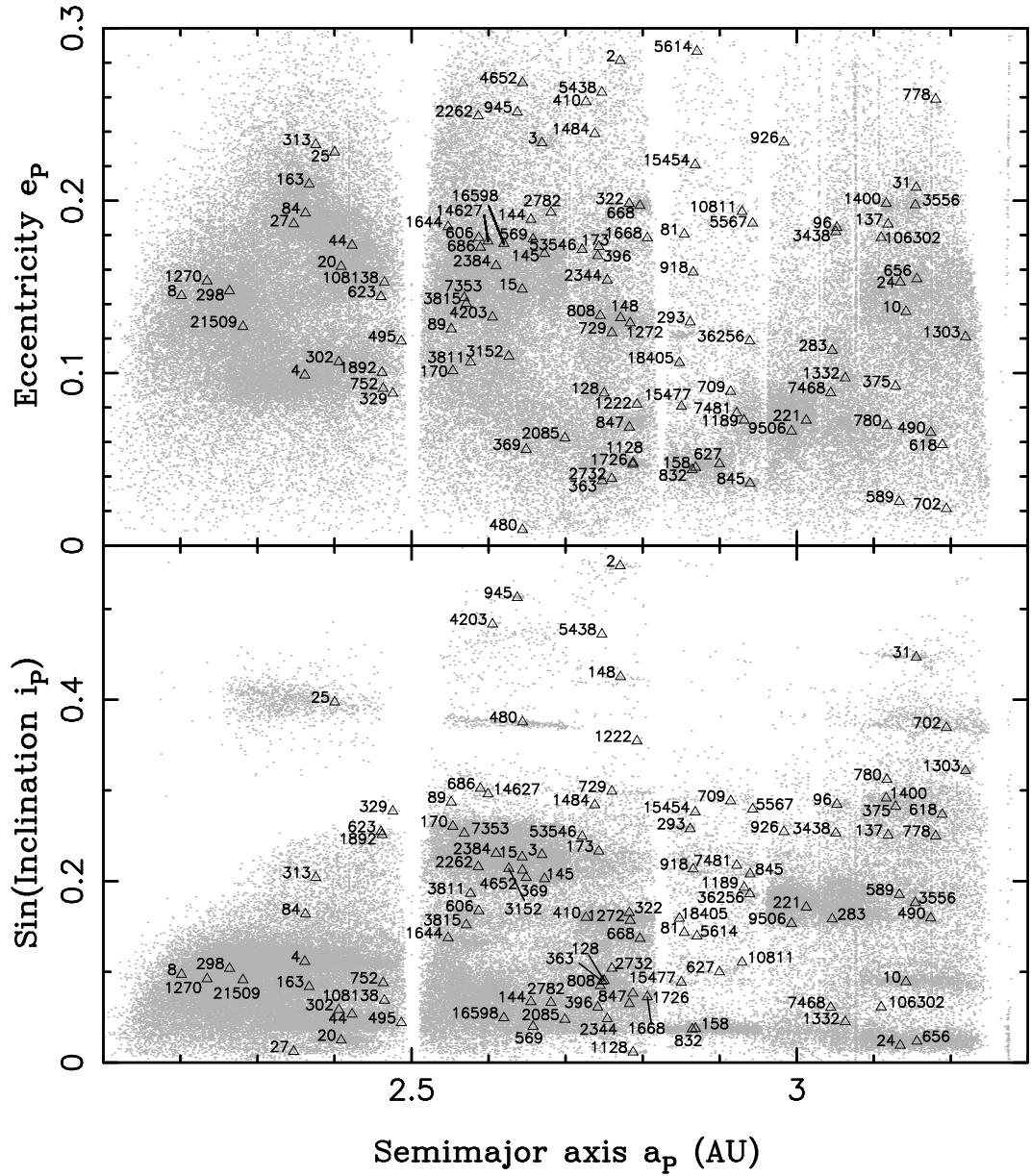}
\caption{\small The orbital location of notable families from Table 2. A triangle is placed at
the orbit of an asteroid after which the family is named. The label near the triangle shows the designation 
number of that asteroid.}   
\label{map}
\end{figure*}

\section{\textbf{SYNTHESIS}}
\smallskip
\noindent
\textbf{7.1 METHOD, FAMILY NAMES AND FIN}
\bigskip

Here we present a synthesis of asteroid families extracted from many past publications (Moth\'e-Diniz et al., 2005; 
Nesvorn\'y et al., 2005; Gil-Hutton, 2006; Parker et al., 2008; Nesvorn\'y 2010, 2012; Novakovi\'c et al., 
2011; Bro\v{z} et al., 2013; Masiero et al., 2011, 2013; Carruba et al., 2013a; Milani et al., 2014). 
The synthesis is based on the numerically computed proper elements (AstDyS catalog from August 2014 with 
the proper elements for 384,337 numbered asteroids; the proper elements of 4016 Jupiter Trojans were taken 
from Rozehnal and Bro\v{z} 2013) and the latest releases of data from the SDSS and WISE.

First, we collected a long list of all (they were 228) families reported in these publications. Second, we 
checked to see whether the clustering in proper elements of each reported group is meaningful, and whether 
the group has reasonably homogeneous colors and albedos (see methods in Section 2). The families that passed 
the preliminary tests were scrutinized further to establish the best cutoff value $d_{\rm cut}$ (the values 
of $d_{\rm cut}$ were informed from past publications), apply the V-shape criterion to identify large 
interlopers (section 4), and estimate $t_{\rm age}$ from the envelope-based method (Section 5).
 
We made sure that each identified group with $d_{\rm cut}$ was complete (i.e., not missing any extension 
that would make sense dynamically), and not part of a larger structure. In some cases, a cut in proper element space 
was applied to avoid the problem of chaining. In several cases, following Masiero et al. (2013), we used an 
albedo cutoff, $p_{\rm V}<0.15$, to identify dark families in the inner and central belts (the (84) Klio, 
(313) Chaldaea, (2262) Mitidika families). Finally, we removed all identified families and searched for 
significant groups in the leftover population. This led to the identification of several 
additional families (e.g., the (589) Croatia, (926) Imhilde, (1332) Marconia families in the outer belt).   

A difficulty we had to deal with is that different authors often listed the same family under different 
names. For example, Masiero et al. (2013) renamed the families after their largest fragments (e.g., the (158) 
Koronis family was referred to as the (208) Lacrimosa family). While it is very useful to identify 
the largest fragment, this naming convention creates chaos because some well-established family names disappear 
(Koronis, Maria, Flora, Gefion, Dora, etc.) and new names are used to refer to the same groups (Lacrimosa, Roma, 
Augusta, Gudiachvili, Zhongolovich, etc.). It is also not clear in many cases what the true largest member is 
((208) Lacrimosa is only slightly larger than (158) Koronis, some large objects have peripheral orbits and 
can be interlopers, etc.). 

A simple solution to this problem is to think of the family names as of {\it labels}, used from historical and
other reasons, which do not necessarily reveal what the largest or lowest-numbered asteroid is. Thus,
the Koronis family would remain the Koronis family, regardless of whether (158) Koronis, (208) Lacrimosa or 
some other large asteroid are the actual family members, largest or not. 

Anticipating that some authors will still prefer to rename some families, here we assign each family a unique Family 
Identifier Number (FIN). The FIN goes from 101 to 399 for families identified from the analytic proper elements: 
1XX for the inner belt (2-2.5 AU), 2XX for the middle belt (2.5-2.82 AU), and 3XX for the outer belt 
(2.82-3.7 AU). The families identified from the synthetic proper elements were assigned the FIN from 401 to 999: 4XX and 
7XX for the low and high inclination families in the inner belt, 5XX and 8XX for the low and high inclination families 
in the middle belt, and 6XX and 9XX for the low and high inclination families in the outer belt. The division between
low and high inclinations was taken at $\sin i_{\rm P}=0.3$ ($i_{\rm P}\simeq17.5^\circ$) which coincides with the 
highest inclinations for which the analytic proper elements are available. The FINs of the low-inclination families 
are 'aligned' such that 101 and 401, 102 and 402, 201 and 501, etc., are the same families identified from two
different datasets. The families among Hungarias with $a<2$~AU, Hildas with $a\simeq3.8$~AU, and Jupiter Trojans with 
$a\simeq5.2$~AU are given the FIN between 001 and 099. 

The main idea behind this notation is that even if a family is eventually renamed, its FIN will remain the same. 
Therefore, it should suffice to list the FIN of the renamed family in a new publication to make it clear how the 
new name relates to the designation(s) used in the past. For example, the Koronis family is given the FIN equal to 605 
(synthetic proper elements, outer belt, low $i$, 5th notable family), and will remain 605 even if the name Lacrimosa 
family will eventually be adopted. Since, however, it can be easily checked that the Koronis and Lacrimosa families 
have the same FIN (assuming that authors list the FIN in their publications), it will be crystal clear that we deal 
the same family. 

A potential difficulty with this scheme would arise if families that were previously assigned a FIN are often found 
to be nonexistent (say, because they are later found statistically insignificant or part of a large structure that 
was not evident in the original data). We suggest that such families are deleted and the associated FINs are left void 
(i.e., not assigned to any family) such that there is no source for confusion. Finally, to avoid having many void FINs 
and not to run out of the FINs with the three digit scheme described above, the FIN should be given only to families 
that are reasonably secure. Other, less secure families can be tentatively listed as family candidates and can wait for 
their FIN until the situation becomes clear with the new data. If the number of families in the inner, middle and/or 
outer belts exceeds 99, the 3-digit FIN scheme described here can be easily modified to a 4-digit scheme by adding 0 
at the second decimal digit, such that, e.g., FIN=401 (the Vesta family) becomes FIN=4001. This change can be  
implemented if the need arises. 
 
\bigskip
\noindent
\textbf{7.2 FAMILY LISTS}
\bigskip

There are 122 notable asteroid families reported in Table 2 and 19 family candidates in Note 11. The vast majority of 
families listed in Table 2 have been reported in many past publications and are clear beyond doubt (Vesta, Massalia, Eunomia, 
Gefion, Koronis, Eos, Themis, etc.). Many additional families, reported in only one or two previous publications, are also 
clear beyond doubt with the new data. Eleven families are reported here for the first time ((329) Svea and (108138) 2001 
GB11 in the inner belt, (89) Julia, (369) Aeria, (1484) Postrema, (2782) Leonidas and (3152) Jones in the central belt, 
(589) Croatia, (926) Imhilde, (1332) Marconia and (106302) 2000 UJ87 in the outer belt). There are 20 notable families in 
the inner belt (the Nysa-Polana complex is counted as 3 families here), 47 in the central belt, 46 in the outer belt, 2 among 
Hildas in the 3:2 resonance with Jupiter, 1 in Hungarias, and 6 among Jupiter Trojans. Of these, seventeen have high orbital 
inclination ($i_{\rm P}>17.5^\circ$). Figure 9 shows the orbital location of the notable families in the main belt.

The lists of members of the notable families can be downloaded from the PDS node (Note 10). Each list contains the following 
information: (1) the asteroid number, (2) $a_{\rm P}$, (3) $e_{\rm P}$, (4) $\sin i_{\rm P}$, and (5) the absolute magnitude $H$ 
from the Minor Planet Center. Also, for families that have the V-shape envelope in $(a_{\rm P},H)$, we fit $C_0$ to this shape 
(Section 4, column 7 in Table 2) and report $C_j/C_0$ for each family member in column~6 of the PDS files. The average albedo 
of each family was obtained from the WISE, and is reported in column 9 of Table~2 ($p_{\rm V}$). The taxonomic type of families, 
reported in column~8 of Table 2, was taken from the previous taxonomic classification of families (Cellino et al. 2002) 
or was deduced from the SDSS colors. 

Columns 5 and 6 of Table 2 report the estimated diameter of the largest member, $D_{\rm LM}$, and diameter of a sphere with volume 
equivalent to that of all fragments, $D_{\rm frag}$. $D_{\rm LM}$ was obtained from AKARI, if available, or from WISE, if 
available, or was estimated from $H$ and average $p_{\rm V}$. The largest member and suspected interlopers with 
$|C_j/C_0|>2$ were excluded in the estimate of $D_{\rm frag}$. The comparison of $D_{\rm LM}$ and $D_{\rm frag}$ helps to establish 
whether a particular breakup event was catastrophic ($D_{\rm frag}>D_{\rm LM}$) or cratering ($D_{\rm frag}<D_{\rm LM}$), but 
note that this interpretation may depend on sometimes uncertain membership of the largest family objects. Also, 
the diameter of the parent body of each family can be estimated as $D_{\rm PB}=(D_{\rm LM}^3+D_{\rm frag}^3)^{1/3}$,
but note that this estimate ignores the contribution of small (unobserved) fragments. 

\bigskip
\noindent
\textbf{7.3 COMPARISON WITH PREVIOUS DATASETS}
\bigskip

The family synthesis presented here is consistent with the results reported in Nesvorn\'y (2012), Bro\v{z} et al. (2013) and 
Carruba et al. (2013a). For example, all families reported in Nesvorn\'y (2012) were found to be real here (except the (46) 
Hestia family that was moved to candidates; see Note 11). Nesvorn\'y (2012), however, used very conservative criteria for 
the statistical significance of a family, and reported only 76 families (or 78 if the Nysa-Polana complex is counted, as it should, 
as three families). Using the 2014 catalog of proper elements, albedo information from Masiero et al. (2013), and validating 
several new families from Milani et al. (2014), here we collect 44 families that did not appear in Nesvorn\'y (2012).
Almost all families reported in Bro\v{z} et al. (2013) also appear here (a notable exception is a large group 
surrounding (1044) Teutonia, which we believe is not a real family; see Note 11), but many new cases were added.
The correspondence with Carruba et al. (2013a) is also good. 

Parker et al. (2008) used $N_{\rm min}=100$ and therefore missed many small families which did not have more than 100
members in the 2008 catalog. Also, given that they used a subset of asteroids with the SDSS colors, even relatively
large families were unnoticed in this work (e.g. the (752) Sulamitis family in the inner belt now has 303 members). The 
high-$i$ families were not reported in Parker et al. (2008), because they only used the analytic proper elements which 
are not available for the high-$i$ orbits. The strength of Parker et al.'s identification scheme was the reliability. 
Indeed, of all families reported in Parker et al. (2008) only the (1044) Teutonia, (1296) Andree and (2007) McCuskey 
(part of the Nysa-Polana complex) are not included among the notable families here (Parker et al.'s (110) Lydia family 
appears here as the (363) Padua family). 

Masiero et al. (2013; hereafter M13) reported 28 new cases and found that 24 old families were lost when 
compared to the family lists in Nesvorn\'y (2012). Most families not listed in M13 are well-defined families 
such as Karin, Beagle, Datura, Emilkowalski, Lucascavin, etc. These families were not listed in M13, because 
they overlap with larger families (and were included in their membership lists in M13) or because they only have 
a few known members (i.e., fall below $N_{\rm min}$ used in M13). On the other hand, we verified that many
new cases reported in Masiero et al. (2013) are genuine new families that can be conveniently found with the albedo 
cutoff (e.g., the (84) Klio, (144) Vibilia, (313) Chaldaea ($=$ (1715) Salli in M13), (322) Phaeo, (623) Chimarea, 
(816) Juliana, and (1668) Hanna families).  These families were included here. In some cases, we found that M13's 
new family barely stands out from the background and it thus seems uncertain. To stay on the safe side, we therefore report 
these cases as the candidate families in Note 11. These families may be real, but their statistical significance 
needs to be carefully tested with the new data. 

Finally, we compare the family synthesis with Milani et al. (2014; hereafter M14), who used
the newest catalogs from all previous works discussed here. They identified many new families which are certain 
beyond doubt (e.g., the (96) Aegle, (618) Elfriede, (2344) Xizang, (3438) Inarradas, (3811) Karma, (7468) Anfimov,
(53546) 2000 BY6; the (96) Aegle and (618) Elfriede families were also reported in M13). These cases highlight 
the strength of the M14 identification scheme and are included in the family synthesis in Table 2. Some 
smaller families located inside bigger families (e.g., the (832) Karin family families in the (158) Koronis family, (656) Beagle 
in (24) Themis) were not reported in M14. In addition, several families were not reported, presumably because the 
QRL was set too low to detect them. This happens, most notably, in the 2.82-2.96~AU region (i.e., between 
the 5:2 and 7:3 resonances), where the number density of asteroids is relatively low (see Figures 1 and 9). The notable cases in 
this region include the (81) Terpsichore, (709) Fringilia, (5567) Durisen, (5614) Yakovlev, (7481) San Marcello and (10811) Lau 
families. A possible solution to this issue would be, in terms of the M14 identification scheme, to use a separate QRL level for the 
2.82-2.96 AU region. Also, several families were split in M14 into several parts. This affects the following families: 
(8) Flora (split in 4 parts), (31) Euphrosyne (3 parts), (221) Eos (5 parts), (702) Alauda (4 parts), (1400) Tirela 
(8 parts), (2085) Henan (4 parts), (4203) Brucato (4 parts) (here we only list families that were split to three or 
more parts in M14).   
\section{\textbf{CONCLUSIONS}}
It is clear from several different arguments that the list of known families must be largely incomplete.  
For example, most families with an estimated parent body size below $\simeq$100-200 km are found to have ages $t_{\rm age}\lesssim 1$ g.y.
(e.g., Bro\v{z} et al., 2013). In contrast, the rate of impacts in the main belt, and therefore the rate of family-forming 
events, should have been roughly unchanging with time over the past $\simeq$3.5 g.y. (and probably raised quite a bit toward the
earliest epochs). So, there must be many missing families with the formation ages $t_{\rm age}>1$ g.y. (Note 12). These 
families are difficult to spot today, probably because they have been dispersed by dynamical processes, lost members by collisional 
grinding, and therefore do not stand out sufficiently well above the dense main-belt background (they are now part of the 
background). 

The significant incompleteness of known families is also indicated by the extrapolation of the number of families 
detected in the 2.82-2.96~AU zone to the whole main belt. As we hinted on at the end of the last section, the 2.82-2.96~AU zone 
is sparsely populated 
such that asteroid families can be more easily detected there. Nearly 20 families with $i_{\rm P}<17.5^\circ$ were 
identified in this region. In comparison, the 2.96-3.3~AU zone, where also $\simeq$20 families with $i_{\rm P}<17.5^\circ$
can be found, is about twice as wide as the 2.82-2.96~AU zone and contains about twice as many large asteroids. 
A straightforward conclusion that can be inferred from this comparison, assuming everything else being equal, is that 
the families in the 2.96-3.3~AU zone are (at least) a factor of $\sim$2 incomplete. A similar argument applies to the 
inner and central belts.  

This is actually good news for future generations of planetary scientists, because this field is open for new discoveries. Figuring
out how to find the missing asteroid families with $t_{\rm age} > 1$ g.y. will not be easy. One way forward would be to improve 
our capability to model the dynamical evolution of main belt asteroids over g.y. timescales, such that we can rewind the clock 
and track fragments back to their original orbits. The modeling of the Yarkovsky effect could be improved, for example, if we 
knew the spin states, densities, conductivity, etc.,  of small main-belt asteroids on individual basis. Another approach would 
consist in identifying families based on the physical properties of their members. While this method is already in use, mainly 
thanks to data from the SDSS and WISE, we anticipate that it can be pushed much further, say, with automated spectroscopic surveys, 
or, in more distant future, with routine sampling missions.
\section{\textbf{NOTES}}
{\footnotesize

\noindent
{\bf Note 1} -- Alternatively, one can use the frequencies $n$, $g$ and $s$ (Carruba and Michtchenko, 
2007, 2009), where $n$ is the mean orbital frequency and $g$ and $s$ are the (constant) precession frequencies 
of the proper perihelion longitude $\varpi_{\rm P}$ and the proper nodal longitude $\Omega_{\rm P}$, respectively. 
The use of frequencies, instead of the proper elements, can be helpful for asteroid families 
near or inside the secular orbital resonances (e.g. the Tina family in the $\nu_6$ resonance; Carruba and Morbidelli, 
2011). 

\noindent
{\bf Note 2} -- Additional methods were developed and/or adapted 
for specific populations of asteroids such as the ones on the high-inclination and high-eccentricity 
orbits (Lema\^{\i}tre and Morbidelli, 1994) or in orbital resonances (Morbidelli, 1993; Milani, 
1993; Beaug\'e and Roig, 2001; Bro\v{z} and Vokrouhlick\'y, 2008; Bro\v{z} and Rozehnal, 2011).

\noindent
{\bf Note 3} -- {\tt http://hamilton.dm.unipi.it/astdys/}

\noindent
{\bf Note 4} -- See chapter 
Bendjoya and Zappal\`a (2002) in the Asteroids III book for a discussion of other clustering 
algorithms such as the Wavelet Analysis Method (WAM) and D-criterion. The WAM was shown to produce 
results that are in a good agreement with those obtained from the HCM (Zappal\`a et al., 1994). 
The D-criterion was originally developed to identify meteorite streams (Southwork and Hawkins, 1963). 
These methods have not been used for the classification of asteroid families in the past decade, 
and we therefore do not discuss them here.

\noindent
{\bf Note 5} -- The SDSS measured flux densities in five bands with effective wavelengths 
354, 476, 628, 769, and 925 nm. The WISE mission measured fluxes in four wavelengths (3.4, 4.6, 12 
and 22~$\mu$m), and combined the measurements with a thermal model to calculate albedos 
($p_V$) and diameters ($D$). The latest public releases of these catalogs include color or albedo data for 
over 100,000 main belt asteroids with known orbits, of which about 25,000 have {\it both} the color and 
albedo measurements. The catalogs are available at\\ 
{\tt www.sdss.org/dr6/products/value\_added/index.html} and  
{\tt irsa.ipac.caltech.edu/Missions/wise.html}.

\noindent
{\bf Note 6} -- Most but not all asteroid 
families are physically homogeneous. The Eos family has the highest diversity of taxonomic classes of any 
known family (e.g., Moth\'e-Diniz et al., 2008). This diversity has led to the suggestion that the Eos parent 
body was partially differentiated. It can also be the source of carbonaceous chondrites (Clark et al., 2009). 
The Eunomia family may be another case of a relatively heterogeneous family (e.g., Nathues et al., 2005). See 
Weiss and Elkins-Tanton (2013) for a review.

\noindent
{\bf Note 7} -- Dell'Oro et al. (2004) attempted to model observed family shapes by Gaussian ellipsoids.
The distribution of $|f|$ obtained in this work was strongly peaked near $\pi/2$, while a more uniform 
distribution between 0 and $\pi$ would be expected if different breakups occurred at random orbital phases.
This result was attributed to the Yarkovsky effect.

\noindent
{\bf Note 8} -- Nesvorn\'y et al. (2008a) found
evidence for a large population of V-type asteroids with slightly lower orbital inclinations 
($i_{\rm P}=3$-4$^\circ$) than the Vesta family ($i_{\rm P}\simeq5^\circ$). Because these asteroids could 
not have dynamically evolved from the Vesta family region to their present orbits in $\sim$1 g.y., they 
are presumably fragments excavated from (4) Vesta's basaltic crust by an earlier impact.

\noindent
{\bf Note 9} -- Other asteroid families whose long-term 
dynamics has been studied in detail, listed here in the alphabetical order, are the 
Adeona family (affected in $e_{\rm P}$ and $i_{\rm P}$ by the 8:3 resonance at 2.705~AU, Carruba et al., 2003),  
Agnia family (inside the $z_1$ resonance; Vokrouhlick\'y et al., 2006c), 
Astrid family (near the border of the 5:2 resonance; Vokrouhlick\'y et al., 2006b),
Eunomia family (Carruba et al., 2007b),
Euphrosyne family (located in a region with many resonances, including $g-g_6=0$, 
near the inner border of the 2:1 resonance; Carruba et al., 2014),  
Erigone family (cut in the middle by the $z_2$ resonance; Vokrouhlick\'y et al., 2006b),
Gefion family (affected by various resonances near 2.75 AU, Carruba et al., 2003, Nesvorn\'y et al., 2009),
Hilda and Schubart families in the 3:2 resonance with Jupiter (Bro\v{z} and Vokrouhlick\'y, 2008),
Hungaria family (perturbed by $2g - g_5 - g_6=0$ and other secular resonances below 1.93 AU;
Warner et al. (2009), Milani et al. (2010), also see Galiazzo et al. (2013, 2014) for the 
contribution of Hungarias to the E-type NEAs and \'Cuk et al. (2014) for their suggested relation 
to the aubrite meteorites), 
Hygiea family (Carruba, 2013; Carruba et al., 2014), 
Massalia family (the part with $a_{\rm P}>2.42$ AU disturbed by the 1:2 resonance with Mars,
Vokrouhlick\'y et al., 2006b),
Merxia family (spread by the 3J-1S-1 three-body resonance at $a_{\rm P}=2.75$ AU; Vokrouhlick\'y et al., 2006b),
Padua family (Carruba, 2009a),
Pallas family (Carruba et al., 2011), 
Phocaea family (Carruba, 2009b), 
Sylvia family ((87) Sylvia has two satellites, possibly related to the impact that produced
the Sylvia family, Vokrouhlick\'y et al., 2010),
and Tina family (Carruba and Morbidelli, 2011).

\noindent
{\bf Note 10} -- {\tt http://sbn.psi.edu/pds/resource/nesvornyfa\\m.html}

\noindent
{\bf Note 11} -- The candidate families are: (929) Algunde, (1296) Andree, (1646) Rosseland, (1942) Jablunka, 
(2007) McCuskey, (2409) Chapman, (4689) Donn, (6246) Komurotoru and (13698) 1998 KF35 in the inner belt, (46) Hestia, 
(539) Palmina, (300163) P/2006 VW 139, (3567) Alvema, (7744) 1986 QA1 in the central belt, and (260) Huberta, (928) Hilrun, (2621)
Goto, (1113) Katja, (8737) Takehiro in the outer belt. We tentatively moved the (46) Hestia family, 
previously known as FIN 503, to the family candidate status, because this group is not convincing with the present 
data. The previously reported groups around (5) Astraea, (1044) Teutonia, (3110) Wagman, (4945) Ikenozenni, 
(7744) 1986 QA1, (8905) Bankakuko, (25315) 1999 AZ8, (28804) 2000 HC81 seem to align with the 
$z_1 = g+s-g_6-s_6=0$ resonance and are probably an artifact of the HCM chaining.   

\noindent
{\bf Note 12} -- The list of known families corresponding to parent bodies with $D>200$ km, on the other hand, is probably reasonably 
complete, because the estimated ages of these families appear to be randomly distributed over 4 g.y. (Bro\v{z} et al., 2013).
These largest families can therefore be used to constrain the collisional history of the asteroid belt (see chapter 
by Bottke~et~al. in this volume).

}

\paragraph{Acknowledgments.} The work 
of MB was supported by the Czech Grant Agency (grant no. P209-12-01308S). The work of VC was supported by the S\~ao Paulo 
State (FAPESP grants no.  2014/06762-2) and Brazilian (CNPq grant no. 305453/2011-4) Grant Agencies.
\bigskip

\centerline\textbf{REFERENCES}
\bigskip
\parskip=0pt
{\small
\baselineskip=11pt

\refs Beaug\'e, C. and Roig, F. (2001) A semi-analytical model for the
motion of the Trojan asteroids: proper elements and families. {\sl
Icarus, 153}, 391--415.

\refs Bendjoya, Ph. and Zappal\`{a}, V. (2002) Asteroid Family Identification.
In: {\sl Asteroids III} (W. Bottke, A. Cellino, P. Paolicchi and 
R.P. Binzel, Eds.), Univ. Arizona Press and LPI, 613--618.

\refs Benz, W. and Asphaug, E. (1999) Catastrophic Disruptions Revisited.
{\sl Icarus, 142}, 5--20.

\refs Bottke, W. F., Vokrouhlick\'{y}, D., Bro\v{z}, M. et al. (2001)
 Dynamical Spreading of Asteroid Families by the Yarkovsky Effect.
{\sl Science, 294}, 1693--1696.

\refs Bottke, W. F., Vokrouhlick\'{y}, D., Rubincam, D. P., and Bro\v{z}, 
M. (2002) The Effect of Yarkovsky Thermal Forces on the Dynamical 
Evolution of Asteroids and Meteoroids.
In: {\sl Asteroids III} (W. Bottke, A. Cellino, P. Paolicchi and 
R.P. Binzel, Eds.), Univ. Arizona Press and LPI, 395--408.

\refs Bottke, W. F., Durda, D. D., Nesvorn\'{y}, D. et al. (2005a)
The fossilized size distribution of the main asteroid belt. 
{\sl Icarus, 175}, 111--140.

\refs Bottke, W. F., Durda, D. D., Nesvorn\'{y}, D. et al. (2005b)
Linking the collisional history of the main asteroid belt to its 
dynamical excitation and depletion.  {\sl Icarus, 179}, 63--94.

\refs Bottke, W. F., Vokrouhlick\'{y}, D., Rubincam, D. P., and Nesvorn\'{y}, 
D. (2006) The Yarkovsky and Yorp Effects: Implications for Asteroid Dynamics.
{\sl Annual Review of Earth and Planetary Sciences,  34}, 157--191.

\refs Bottke, W. F., Vokrouhlick\'{y}, D., and Nesvorn\'{y}, D. (2007) 
An asteroid breakup 160 Myr ago as the probable source of the K/T impactor.
{\sl Nature,  449}, 48--53. 

\refs Bottke, W.~F., 
Vokrouhlick{\'y}, D., Walsh, K.~J., Delbo, M., Michel, P., Lauretta, D.~S., 
Campins, H., Connolly, H.~C., Scheeres, D.~J., and Chelsey, S.~R.\ (2015) In 
search of the source of asteroid (101955) Bennu: Applications of the 
stochastic YORP model.\ {\it Icarus, 247}, 191--217. 

\refs Bottke, W. F. et al. (2015) Collisional evolution of the asteroid belt.
In : {\sl Asteroids IV} (P. Michel, F. E. DeMeo, W. Bottke Eds.),
Univ. Arizona Press and LPI.

\refs Bowell, E., Muinonen, K., and Wasserman, L. H. (1994) 
A Public-Domain Asteroid Orbit Data Base. Asteroids, Comets, Meteors 1993, 160, 477. 

\refs Bro\v{z}, M. and Morbidelli, A. (2013) The Eos family halo.
{\sl Icarus,  223}, 844--849.

\refs Bro\v{z}, M. and Vokrouhlick\'{y}, D. (2008) Asteroid families 
in the first-order resonances with Jupiter.  {\it MNRAS, 390}, 715--732.

\refs Bro\v{z}, M., Vokrouhlick\'{y}, D., Morbidelli, A., 
Nesvorn\'{y}, D., and Bottke, W. F. (2011) Did the Hilda collisional family 
form during the late heavy bombardment? {\it MNRAS, 414}, 2716--2727.

\refs Bro\v{z}, M. and Rozehnal, J. (2011) Eurybates - the only asteroid 
family among Trojans? {\it MNRAS, 414}, 565--574.
	
\refs Bro\v{z}, M., Morbidelli, A.,  Bottke, W. F., et al. (2013)
Constraining the cometary flux through the asteroid belt during the 
late heavy bombardment.  {\sl Astron. Astrophys.,  551}, A117.

\refs Brunetto, R. et al. (2015) Asteroid Surface Alteration by Space 
Weathering Processes. In : {\sl Asteroids IV} (P. Michel, F. E. 
DeMeo, W. Bottke Eds.), Univ. Arizona Press and LPI.

\refs Campbell, M. (1995) Golden Eye, MGM/UA Distribution Company

\refs Carruba, V. (2009a) The (not so) peculiar case of the Padua family. 
{\it MNRAS, 395}, 358--377.

\refs Carruba, V., (2009b) An analysis of the region of the Phocaea 
dynamical group.  {\it MNRAS, 398}, 1512--1526.

\refs Carruba, V., (2010) The stable archipelago in the region of the 
Pallas and Hansa families. {\it MNRAS, 408}, 580--600.

\refs Carruba, V. and Michtchenko, T.A. (2007) A frequency approach to
identifying asteroid families. {\sl Astron. Astrophys.,  75},
1145--1158.

\refs Carruba, V. and Michtchenko, T. A. (2009)  A frequency approach 
to identifying asteroid families.  II. Families interacting 
with non-linear secular resonances and low-order mean-motion 
resonances.  {\sl Astron. Astrophys.,  493}, 267--282.

\refs Carruba, V. and Morbidelli A. (2011)  On the first nu(6) 
anti-aligned librating asteroid family of Tina.  
{\it MNRAS, 412}, 2040--2051.

\refs Carruba, V., Burns, J. A., Bottke, W. F., and Nesvorn\'{y} D. (2003) 
Orbital evolution of the Gefion and Adeona asteroid families: 
close encounters with massive asteroids and the Yarkovsky effect.  
{\sl Icarus,  162}, 308--327.

\refs Carruba, V., Michtchenko, T. A., Roig, F., Ferraz-Mello, S., and 
Nesvorn\'{y}, D. (2005)  On the V-type asteroids outside the Vesta family.  
I.  Interplay of nonlinear secular resonances and the Yarkovsky effect: 
the cases of 956 Elisa and 809 Lundia.  {\sl Astron. Astrophys.,  441}, 
819--829.

\refs Carruba, V., Roig, F., Michtchenko, T. A., Ferraz-Mello, S., and 
Nesvorn\'{y} D. (2007a) Modeling close encounters with massive 
asteroids: a Markovian approach. An application to the Vesta family.  
{\sl Astron. Astrophys.,  465}, 315--330.

\refs Carruba, V., Michtchenko, T. A., and Lazzaro D. (2007b)  
On the V-type asteroids outside the Vesta family.  II.  
Is (21238) 1995 WV7 a fragment of the long-lost basaltic crust of 
(15) Eunomia? {\sl Astron. and Astrophys.,  473}, 967--978.

\refs Carruba, V., Machuca, 
J.~F., and Gasparino, H.~P.\ (2011) Dynamical erosion of asteroid groups in the 
region of the Pallas family.\ {\it MNRAS, 412}, 2052--2062. 

\refs Carruba, V., Huaman, M. E., Douwens, S., and Domingos, R. C. (2012) 
Chaotic diffusion caused by close encounters with several massive 
asteroids. {\sl Astron. Astrophys.,  543}, A105.

\refs Carruba, V., Domingos, R. C., Nesvorn\'{y}, D., et al. (2013a) 
A multi-domain approach to asteroid families 
identification.  {\it MNRAS, 433}, 2075--2096.

\refs Carruba, V., Huaman, M. E., Domingos, R. C., and Roig F. (2013b) 
Chaotic diffusion caused by close encounters with several massive 
asteroids II: the regions of (10) Hygiea, (2) Pallas, and (31) Euphrosyne, 
{\sl Astron. Astrophs.,  550}, A85.

\refs Carruba, V., Aljbaae, S., and Souami D. (2014), Peculiar Euphrosyne, 
{\sl Astrophys. J.,  792}, 46--61.

\refs Cellino, A., Bus, S. J., Doressoundiram, A. and Lazzaro, D. (2002)
 Spectroscopic Properties of Asteroid Families. In: {\sl Asteroids III}
(W. Bottke, A. Cellino, P. Paolicchi and R.P. Binzel, Eds.),
Univ. Arizona Press and LPI, 633--643.

\refs Cellino, A., Dell'Oro, A., and Zappal\`a, V. (2004) Asteroid families: 
open problems.  {\sl Planetary and Space Science,  52}, 1075--1086.

\refs Cellino, A., Dell'Oro, A., and Tedesco, E. F. (2009) Asteroid families: 
Current situation.  {\sl Planetary and Space Science,  57}, 173--182.

\refs Cellino, A., Bagnulo, S., Tanga, P., Novakovi\'{c}, B., and Delb\'{o}, 
M. (2014)   A successful search for hidden Barbarians in the Watsonia 
asteroid family. {\it MNRAS Letters, 439}, 75--79.

\refs Cibulkov\'{a}, H., Bro\v{z}, M., and Benavidez, P. G. (2014) A six-part 
collisional model of the main asteroid belt. {\sl Icarus, 241}, 358--372.

\refs Cimrman, J. (1917) On the literal expansion of the perturbation 
Hamiltonian and its applicability to the curious assemblages of minor planets. 
{\sl The Cimrman Bulletin,  57}, 132--135.

\refs Clark, B.~E., Ockert-Bell, M.~E., Cloutis, E.~A., Nesvorn\'y, D., Moth{\'e}-Diniz, T., 
and Bus, S.~J.\ (2009)\ Spectroscopy of K-complex asteroids: Parent bodies of 
carbonaceous meteorites?\ {\sl Icarus,  202}, 119--133. 
	
\refs \'{C}uk, M. Gladman, B. J., and Nesvorn\'{y}, D. (2014) 	
Hungaria asteroid family as the source of aubrite meteorites.
{\sl Icarus,  239}, 154--159.

\refs Delisle, J.-B. and Laskar, J. (2012) Chaotic diffusion of the Vesta 
family induced by close encounters with massive asteroids.  
{\sl Astron. Astrophys.,  540}, A118.

\refs Dell'Oro, A., Bigongiari, G., Paolicchi, P., and  Cellino, A. (2004)
Asteroid families: evidence of ageing of the proper elements. 
{\sl Icarus,  169}, 341--356.

\refs Dell'Oro, A. and Cellino, A. (2007) The random walk of Main Belt 
asteroids: orbital mobility by non-destructive collisions. {\it MNRAS, 380}, 
399--416.

\refs Dermott, S.~F., 
Nicholson, P.~D., Burns, J.~A., Houck, J.~R. (1984) Origin of the solar 
system dust bands discovered by IRAS.\ {\it Nature, 312}, 505--509. 

\refs Dermott, S.~F., Kehoe, 
T.~J.~J., Durda, D.~D., Grogan, K., Nesvorn{\'y}, D.\ (2002) Recent 
rubble-pile origin of asteroidal solar system dust bands and asteroidal 
interplanetary dust particles.\ {\it Asteroids, Comets, and Meteors, 
500}, 319--322. 

\refs Durda, D.~D., Bottke, 
W.~F., Enke, B.~L., Merline, W.~J., Asphaug, E., Richardson, D.~C., 
and Leinhardt, Z.~M.\ (2004) The formation of asteroid satellites in large 
impacts: results from numerical simulations.\ {\it Icarus, 170}, 243--257. 

\refs Durda, D. D., Bottke, W. F., Nesvorn\'{y}, D. et al. (2007) 
Size-frequency distributions of fragments from SPH/ N-body simulations 
of asteroid impacts: Comparison with observed asteroid families. 
{\sl Icarus,  186}, 498--516.

\refs Dykhuis, M. and Greenberg, R. (2015) Collisional family structure 
within the Nysa-Polana complex.\ {\sl Icarus}, ArXiv e-prints arXiv:1501.04649. 

\refs Dykhuis, M.~J., Molnar, L., Van Kooten, S.~J., Greenberg, R.\ (2014) 
Defining the Flora Family: Orbital properties, reflectance properties and age.
{\sl Icarus,  243}, 111--128. 

\refs Farley, K. A., Vokrouhlick\'{y}, D., Bottke, W. F., and Nesvorn\'{y}, D.
(2006)  A late Miocene dust shower from the break-up of an 
asteroid in the main belt.  {\sl Nature,  439}, 295--297.

\refs Florczak, M., Lazzaro, 
D., and Duffard, R. (2002) Discovering New V-Type Asteroids in the Vicinity of 
4 Vesta.\ {\it Icarus, 159}, 178--182. 

\refs Fujiwara, A., Cerroni, P., Davis, D., Ryan, E., and di Martino, M.\ (1989)\ 
Experiments and scaling laws for catastrophic collisions.\ {\sl Asteroids II}, 
240--265. 

\refs Galiazzo, M. A., Bazs\'{o}, \'{A}., and Dvorak, R. (2013) 
Fugitives from the Hungaria region: Close encounters and 
impacts with terrestrial planets. 
{\sl Planet. and Space Sci.,  84}, 5--13.

\refs Galiazzo, M. A., Bazs\'{o}, \'{A}., and Dvorak, R. (2014)
The Hungaria Asteroids: close encounters and impacts with terrestrial planets.
Memorie della Societa Astronomica Italiana Supplement, v.26, p.38.

\refs Gil-Hutton, R. (2006) Identification of families among highly 
inclined asteroids.  {\sl Icarus,  183}, 93--100.

\refs Harris, A.~W., Mueller, 
M., Lisse, C.~M., and Cheng, A.~F.\ (2009)\ A survey of Karin cluster asteroids 
with the Spitzer Space Telescope.\ {\sl Icarus,  199}, 86--96. 

\refs Hirayama, K. (1918) Groups of asteroids probably of common origin.
{\sl Astron. J.,  31}, 185--188.

\refs Ivezi\'{c}, \v{Z}., Tabachnik, S., Rafikov, R. et al. (2001) 
Solar System Objects Observed in the Sloan Digital Sky Survey Commissioning 
Data. {\sl Astron. J.  122}, 2749--2784.

\refs Jenniskens, P. (2015) 
In  {\sl Asteroids IV} (P. Michel, F. E. DeMeo, W. Bottke Eds.),
Univ. Arizona Press and LPI.

\refs Jewitt, D. (2015) The Activated Asteroids.  In  
{\sl Asteroids IV} (P. Michel, F. E. DeMeo, W. Bottke Eds.),
Univ. Arizona Press and LPI.

\refs Kne\v zevi\'c, Z., Milani, A., Farinella, P., Froeschle, Ch., 
and Froeschle, Cl. (1991) Secular resonances from 2 to 50 AU. 
{\sl Icarus,  93}, 316--330. 

\refs Kne\v zevi\'c, Z. and Pavlovi\'{c}, R., (2000)  Young Age for 
the Veritas Asteroid Family Confirmed?  {\it Earth, Moon, and Planets,  
88}, 155--166.

\refs Kne\v zevi\'c, Z. and Milani, A. (2000) Synthetic proper elements
for outer main belt asteroids. {\sl Celest. Mech. Dyn. Astron.,  78},
17--46.

\refs Kne\v zevi\'c Z., Lema\^{\i}tre, A. and Milani, A. (2002) The
determination of asteroid proper elements. In {\sl Asteroids III}
(W. Bottke, A. Cellino, P. Paolicchi and R.P. Binzel, Eds.),
Univ. Arizona Press and LPI, 603--612.

\refs Lema\^{\i}tre A. and Morbidelli A. (1994) Proper elements for highly
inclined asteroidal orbits. {\sl Cel. Mech. Dyn. Astron., 60},
29--56.

\refs Low, F. J., Young, E., Beintema, D. A., Gautier, T. N., et al. (1984)
Infrared cirrus - New components of the extended infrared emission.  
{\sl Astrophys. J.,  278}, 19--22.

\refs Mainzer, A., Grav, T., Masiero, J., et al. (2011)  NEOWISE Studies 
of Spectrophotometrically Classified Asteroids: Preliminary Results. 
{\sl Astrophys. J.,  741}, 90--115.

\refs Marchi, S., McSween, H. Y., O'Brien, D. P., et al. (2012) The Violent 
Collisional History of Asteroid 4 Vesta.  {\sl Science,  336}, 690.

\refs Marchis, F., and 11 
colleagues 2014.\ The Puzzling Mutual Orbit of the Binary Trojan Asteroid 
(624) Hektor.\ {\it Astrophys. J., 783}, LL37. 

\refs Marsden, B. G. (1980) The Minor Planet Center.  {\sl Cel. Mech. 
Dyn. Astron., 22}, 63--71.
	
\refs Marzari, F., Farinella, P., and Davis, D. R. (1999) 
Origin, Aging, and Death of Asteroid Families.
{\sl Icarus,  142}, 63--77.
	
\refs Masiero, J. R., Mainzer, A. K., Grav, T., et al. (2011)
 Main Belt Asteroids with WISE/NEOWISE. I. Preliminary Albedos and Diameters.
{\sl Astrophys. J.,  741}, 68--88.

\refs Masiero, J. R., Mainzer, A. K., Grav, T., Bauer, J. M., and Jedicke, R.
(2012) Revising the Age for the Baptistina Asteroid Family Using 
WISE/NEOWISE Data.  {\sl Astrophys. J.,  759}, 14--28.

\refs Masiero, J. R., Mainzer, A. K., Bauer, J. M., et al. (2013) 
Asteroid Family Identification Using the Hierarchical Clustering 
Method and WISE/NEOWISE Physical Properties. {\sl Astrophys. J.,  770}, 
7--29.

\refs Masiero, J. R. et al. (2015) Physical Properties of Asteroid Families, 
In {\sl Asteroids IV} (P. Michel, F. E. DeMeo, W. Bottke Eds.),
Univ. Arizona Press and LPI.

\refs Michel, P., Benz, W., Tanga, P.,  and Richardson, D. C. (2001)
Collisions and Gravitational Reaccumulation: Forming Asteroid 
Families and Satellites.  {\sl Science,  294}, 1696--1700. 

\refs Michel, P., Benz, W., and Richardson, D. C. (2003) Disruption of 
fragmented parent bodies as the origin of asteroid families.  
{\it Nature, 421}, 608--611.

\refs Michel, P.,  Benz, W., and Richardson, D. C., (2004)
Catastrophic disruption of pre-shattered parent bodies.
{\sl Icarus,  168}, 420--432.

\refs Michel, P., Jutzi, M., Richardson, D.~C., and Benz, W. (2011) 
The Asteroid Veritas: An intruder in a family named after it?
{\sl Icarus,  211}, 535--545. 

\refs Michel P. et al. (2015) Collisional Formation and Modeling of 
Families. In : {\sl Asteroids IV} (P. Michel, F. E. DeMeo, W. Bottke Eds.),
Univ. Arizona Press and LPI.

\refs Migliorini, F., 
Zappal{\`a}, V., Vio, R., and Cellino, A.\ (1995)\ Interlopers within asteroid 
families.\ {\sl Icarus,  118}, 271--291. 

\refs Milani A. (1993) the Trojan asteroid belt: proper elements, chaos,
stability and families, {\sl Cel. Mech. Dyn. Astron., 57}, 59--94.

\refs Milani, A. and Kne\v zevi\'c, Z. (1990) Secular perturbation
theory and computation of asteroid proper elements. {\sl Celestial
Mechanics,  49}, 347--411.

\refs Milani, A. and Kne\v zevi\'c, Z. (1994) Asteroid proper elements
and the dynamical structure of the asteroid main belt. {\sl Icarus 
107}, 219--254.
	
\refs Milani, A. and Farinella, P. (1994)  The age of the Veritas 
asteroid family deduced by chaotic chronology.  {\sl Nature,  370}, 40--42.

\refs Milani, A., Kne\v{z}evi\'{c}, Z., Novakovi\'{c}, B., 
and Cellino, A. (2010) Dynamics of the Hungaria asteroids.  
{\sl Icarus,  207}, 769--794.

\refs Milani, A., Cellino, A., Kne\v{z}evi\'{c}, Z. et al. (2014) Asteroid 
families classification: Exploiting very large datasets.  {\sl Icarus, 239}, 46--73.

\refs Minton, D. and  Malhotra, R. (2009) A record of planet migration in 
the main asteroid belt. {\sl Nature,  457}, 1109--1111.

\refs Molnar, L. A. and Haegert, M. J. (2009)  Details of Recent 
Collisions of Asteroids 832 Karin and 158 Koronis. AAS/Division 
for Planetary Sciences Meeting Abstracts \#41, 41, \#27.05.  

\refs Morbidelli, A. (1993) Asteroid secular resonant proper elements. 
{\sl Icarus,  105}, 48--66.

\refs Morbidelli, A., 
Brasser, R., Gomes, R., Levison, H.~F., Tsiganis, K.\ (2010) Evidence from 
the Asteroid Belt for a Violent Past Evolution of Jupiter's Orbit.\
{\it Astron. J., 140}, 1391--1401. 

\refs Moth\'{e}-Diniz, T., Roig, F., and Carvano, J. M. (2005) Reanalysis of 
asteroid families structure through visible spectroscopy.  {\sl Icarus, 
 174}, 54--80.

\refs Moth\'{e}-Diniz, T., Carvano, J. M., Bus, S. J., et al. (2008)
Mineralogical analysis of the Eos family from near-infrared spectra.
{\sl Icarus,  195}, 277--294. 

\refs Nathues, A., Mottola, 
S., Kaasalainen, M., and Neukum, G.\ (2005)\ Spectral study of the Eunomia 
asteroid family. I. Eunomia.\ {\sl Icarus,  175}, 452--463. 

\refs Nesvorn\'y, D. (2010) Nesvorny HCM Asteroid Families V1.0. NASA 
Planetary Data System, 133.

\refs Nesvorn\'y, D. (2012) Nesvorny HCM Asteroid Families V2.0. NASA 
Planetary Data System, 189.

\refs Nesvorn\'{y} D. and Morbidelli, A. (1999)  An Analytic Model of 
Three-Body Mean Motion Resonances. {\it Celestial Mechanics and Dynamical 
Astronomy, 71}, 243--271.

\refs Nesvorn\'{y} D. and Bottke, W. F. (2004) Detection of the Yarkovsky 
effect for main-belt asteroids. {\sl Icarus,  170}, 324--342.

\refs Nesvorn\'y, D. and Vokrouhlick\'y, D. (2006) New Candidates for Recent
Asteroid Breakups. {\sl Astron. J.,  132}, 1950--1958.

\refs Nesvorn\'{y} D., Bottke, W. F., Dones, L., and Levison, H. F. (2002a)
The recent breakup of an asteroid in the main-belt region.  {\sl Nature, 
 417}, 720--771.

\refs Nesvorn\'{y}, D., Morbidelli, A., Vokrouhlick\'{y}, D., 
Bottke, W. F., and Bro\v{z}, M. (2002b) The Flora Family: A Case of the 
Dynamically Dispersed Collisional Swarm?  {\sl Icarus,  157}, 
155--172.

\refs Nesvorn\'{y}, D., Ferraz-Mello, S., Holman, M., and Morbidelli, A.
(2002c) Regular and Chaotic Dynamics in the Mean-Motion Resonances: 
Implications for the Structure and Evolution of the Asteroid Belt.
In: {\sl Asteroids III} (W. Bottke, A. Cellino, P. Paolicchi and 
R.P. Binzel, Eds.), Univ. Arizona Press and LPI, 379--394.

\refs Nesvorn\'{y} D., Bottke, W. F., Levison, H. F., and Dones, L. (2003)
Recent Origin of the Solar System Dust Bands.  {\sl Astrophys. J., 
 591}, 486--497.

\refs Nesvorn\'{y} D., Jedicke, R., Whiteley, R. J., and Ivezi\'{c}, \v{Z}. 
(2005)  Evidence for asteroid space weathering from the Sloan Digital 
Sky Survey.  {\sl Icarus,  173}, 132--152.

\refs Nesvorn{\'y}, D., 
Enke, B.~L., Bottke, W.~F., Durda, D.~D., Asphaug, E., and Richardson, D.~C.\ 
(2006a)\ Karin cluster formation by asteroid impact.\ {\sl Icarus,  183}, 296--311. 

\refs Nesvorn{\'y}, D., 
Bottke, W.~F., Vokrouhlick{\'y}, D., Morbidelli, A., and Jedicke, R.\ (2006b)\ 
Asteroid families.\ {\it Asteroids, Comets, Meteors, 229}, 289-299. 

\refs Nesvorn\'y, D., Vokrouhlick\'y, D., and Bottke, W.F. (2006c) The Breakup
of a Main-Belt Asteroid 450 Thousand Years Ago. {\sl Science,  312}, 1490.

\refs Nesvorn\'y, D., Roig, F., Gladman, B, et al. (2008a) 
Fugitives from the Vesta family. {\sl Icarus,  183}, 85--95.

\refs Nesvorn\'{y}, D., Bottke, W. F., Vokrouhlick\'{y}, D. et al. (2008b)
Origin of the Near-Ecliptic Circumsolar Dust Band. {\sl Astrophys. J., 
 679}, 143--146.

\refs Nesvorn\'{y}, D., Vokrouhlick\'{y}, D., Morbidelli, A., 
and Bottke, W. F. (2009) Asteroidal source of L chondrite meteorites.
{\sl Icarus,  200}, 698--701. 

\refs Novakovi\'c, B. (2010) Portrait of Theobalda as a young asteroid family.
{\it MNRAS, 407}, 1477--1486.

\refs Novakovi\'c, B., Tsiganis, K., and Kne\v{z}evi\'c, Z. (2010) Dynamical 
portrait of the Lixiaohua asteroid family. 
{\it Celestial Mechanics and Dynamical Astronomy, 107}, 35--49.

\refs Novakovi\'c, B., Tsiganis, K., and Kne\v{z}evi\'c, Z. (2010) Chaotic 
transport and chronology of complex asteroid families. {\it MNRAS, 402}, 1263--1272.

\refs Novakovi\'c, B., Cellino, A., and Kne\v{z}evi\'c, Z. (2011) Families among
high-inclination asteroids. {\it Icarus, 216}, 69--81.

\refs Novakovi\'c, B., Dell'Oro, A., Cellino, A., and Kne\v{z}evi\'{c}, Z. 
(2012a) Recent collisional jet from a primitive asteroid. 
{\it MNRAS, 425}, 338--346.

\refs Novakovi\'c, B., Hsieh, H. H., and Cellino, A. (2012b)
P/2006 VW139: a main-belt comet born in an asteroid collision?
{\it MNRAS, 424}, 1432--1441.

\refs Novakovi\'c, B. Hsieh, H.H., Cellino, A., Micheli, M., and Pedanim, M.
(2014)
Discovery of a young asteroid cluster associated with P/2012~F5 (Gibbs).
{\sl Icarus,  231}, 300--309.

\refs Parker, A., Ivezi\'{c}, \v{Z}., Juri\'{c}, M., et al. (2008) The 
size distributions of asteroid families in the SDSS Moving Object Catalog 4. 
{\sl Icarus,  198}, 138--155.

\refs Reddy, V., Carvano, J. M., Lazzaro, D. et al. (2011) 
Mineralogical characterization of Baptistina Asteroid Family: 
Implications for K/T impactor source. {\sl Icarus,  216}, 184--197.

\refs Rozehnal, J. and  Bro\v{z}, M. (2013) Jovian Trojans: Orbital 
structures versus the WISE data.  American Astronomical Society, 
DPS meeting \#45, \#112.12.

\refs Ro\v{z}ek, A., Breiter, S., and Jopek, T.~J. (2011) Orbital similarity functions - 
application to asteroid pairs.\ {\sl MNRAS, 412}, 987--994. 

\refs Rubincam, D. P. (1995) Asteroid orbit evolution due to thermal 
drag.  {\sl Journal of Geophys. Research,  100}, 1585--1594.

\refs Rubincam, D. P. (2000) Radiative Spin-up and Spin-down of 
Small Asteroids. {\sl Icarus,  148}, 2--11.

\refs Roig, F., Ribeiro, A.~O., and Gil-Hutton, R.\ (2008)\ 
Taxonomy of asteroid families among the Jupiter Trojans: comparison between 
spectroscopic data and the Sloan Digital Sky Survey colors.\ 
{\sl Astron. Astrophys.,  483}, 911--931. 

\refs Southworth, R. B. and Hawkins, G. S. (1963) Statistics of meteor 
streams.  {\it Smithsonian Contributions to Astrophysics, 7}, 261.

\refs Tsiganis, K., Kne\v{z}evi\'{c}, Z., and Varvoglis, H. (2007)
Reconstructing the orbital history of the Veritas family.  
{\sl Icarus,  186} 484--497.

\refs Usui, F., Kasuga, T., Hasegawa, S., et al. (2013) Albedo 
Properties of Main Belt Asteroids Based on the All-Sky Survey of 
the Infrared Astronomical Satellite AKARI, {\sl Astrophys. J.,  762}, 
56--70.

\refs Vernazza, P., Binzel, R. P., Thomas, C. A., et al. (2008) 
Compositional differences between meteorites and near-Earth asteroids.
{\sl Nature,  454}, 858--860. 

\refs Vokrouhlick{\'y}, D. and Nesvorn{\'y}, D.\ (2008) Pairs of Asteroids Probably 
of a Common Origin.\ {\it Astron. J.,  136}, 280--290. 

\refs Vokrouhlick\'{y}, D., and Nesvorn\'{y}, D. (2011) Half-brothers in 
the Schulhof Family? {\sl Astron. J.,  142}, 26--34.

\refs Vokrouhlick\'{y}, D., Bro\v{z}, M., Morbidelli, A., et al. (2006a)
Yarkovsky footprints in the Eos family.  {\sl Icarus,  182}, 92--117. 

\refs Vokrouhlick\'{y}, D., Bro\v{z}, M., Bottke, W. F., Nesvorn\'{y}, D., 
and Morbidelli, A. (2006b)  Yarkovsky/YORP chronology of asteroid families.
{\sl Icarus,  182}, 118--142.

\refs Vokrouhlick\'{y}, D., Bro\v{z}, M., Bottke, W. F., Nesvorn\'{y}, D., 
and Morbidelli, A. (2006c)  The peculiar case of the Agnia asteroid family
{\sl Icarus,  183}, 349--361.
	
\refs Vokrouhlick\'{y}, D. and Nesvorn\'{y}, D. (2009)  The Common Roots of 
Asteroids (6070) Rheinland and (54827) 2001 NQ8. {\sl Astron. J.,  137}, 
111--117.
	
\refs Vokrouhlick\'{y}, D., Durech, J., Michalowski, T., et al. (2009)
Datura family: the 2009 update. {\sl Astron. Astrophys.,  507}, 495--504.

\refs Vokrouhlick\'{y}, D., Nesvorn\'{y}, D., Bottke, W. F., and Morbidelli, A.
(2010)  Collisionally Born Family About 87 Sylvia. {\sl Astron. J.,  139}, 
2148--2158.

\refs Vokrouhlick\'{y}, D., \v{D}urech, J., Polishook, D., et al. (2011)
Spin Vector and Shape of (6070) Rheinland and Their Implications. 
{\sl Astron. J.,  142}, 159--167.

\refs Vokrouhlick\'{y}, D. et al. (2015) Yarkovsky and YORP. 
In {\sl Asteroids IV} (P. Michel, F. E. DeMeo, W. Bottke Eds.),
Univ. Arizona Press and LPI.
	
\refs Walsh, K. J., Delbo, M., Bottke, W. F., et al. (2013) 
Introducing the Eulalia and new Polana asteroid families: Re-assessing 
primitive asteroid families in the inner Main Belt.  {\sl Icarus,  225},
283--297. 

\refs Warner, B. D., Harris, A. W., Vokrouhlick\'{y}, D., et al. (2009)
Analysis of the Hungaria asteroid population. {\sl Icarus,  204}, 
172--182.

\refs Weiss, B.~P., 
Elkins-Tanton, L.~T.\ (2013)\ Differentiated Planetesimals and the Parent 
Bodies of Chondrites.\ {\sl Annual Review of Earth and Planetary Sciences,  41}, 
529--560. 

\refs Willman, M., Jedicke, 
R., Nesvorn{\'y}, D., Moskovitz, N., Ivezi{\'c}, {\v Z}., and Fevig, R.\ (2008) 
Redetermination of the space weathering rate using spectra of Iannini 
asteroid family members.\ {\it Icarus, 195}, 663--673. 

\refs Wisdom, J. (1982) The origin of the Kirkwood gaps - A mapping for 
asteroidal motion near the 3/1 commensurability.  {\sl Astron. J.,  87}, 
577--593. 

\refs Zappal\`a, V., Cellino, A., Farinella, P., and Kne\v{z}evi\'{c}, Z. (1990)
Asteroid families. I - Identification by hierarchical clustering and 
reliability assessment.  {\sl Astron. J.,  100}, 2030--2046.

\refs Zappal\`a, V., Cellino, A., Farinella, P., and Milani, A. (1994)
Asteroid families. 2: Extension to unnumbered multiopposition asteroids.
{\sl Astron. J., 107}, 772--801.

\refs Zappal\`{a}, V., Cellino, A., dell'Oro, A., Paolicchi, P. (2002)
Physical and Dynamical Properties of Asteroid Families.  In 
{\sl Asteroids III} (W. Bottke, A. Cellino, P. Paolicchi and R.P. 
Binzel, Eds.), Univ. Arizona Press and LPI, 619--631.

}

\clearpage
\setcounter{table}{1}
\begin{sidewaystable}[h!]
\small
\vspace*{10.cm}
\caption{Notable asteroid families. The columns are the: (1) Family Identification Number 
(FIN), (2) family name, (3) cutoff distance ($d_{\rm cut}$; asterisk denotes $d_{\rm cut}$ used on
a subset of asteroids with $p_{\rm V}<0.15$), (4) number of family members identified with 
$d_{\rm cut}$, (5) largest member(s) in the family (either the number designation of the 
largest member(s), in parenthesis, if different from the asteroid after which the family is named, 
or the estimated diameter of the largest member, $D_{\rm LM}$), 
(6) diameter of a sphere with volume equivalent to that of all fragments ($D_{\rm frag}$, (7) $C_0$ 
parameter defined in Section 4, (8) taxonomic type, (9) mean geometric albedo from WISE ($p_{\rm V}$), 
and (10) various references and notes. We do not report $t_{\rm age}$ here but note that $t_{\rm age}$ 
can be estimated from $C_0$ given in column 7 and Eq.~(2). 
See {\tt http://sirrah.troja.mff.cuni.cz/\~{}mira/mp/fams/} for additional information on asteroid 
families (Bro\v{z} et al., 2013).} 
 
\begin{center}  
\begin{tabular}{llrrrrrrrl}
\hline
FIN    & Family  & $d_{\rm cut}$   & \# of  & $D_{\rm LM}$ & $D_{\rm frag}$    & $C_0$        & Tax. & $p_{\rm V}$ & References and Notes     \\
       & Name    &  (m s$^{-1}$)  &  mem.  & (km)         & (km)            & ($10^{-4}$ AU)  & Type &            &            \\    
\hline
\multicolumn{10}{c}{\normalsize \it Hungarias, Hildas and Jupiter Trojans}\\[1.mm]            
001 & 153 Hilda      & 130    & 409        & 164 & -- & --     & C    & 0.04        & Bro\v{z} et al. (2011) \\
002 & 1911 Schubart  & 60     & 352        & 80  & 91 & --     & C    & 0.03        & Bro\v{z} and Vokrouhlick\'y (2008) \\   
003 & 434 Hungaria   & 100    & 2965       & 10  & 24 & $0.3\pm0.1$ & E   & 0.35    & Warner et al. (2010), Milani et al. (2011) \\
004 & 624 Hector     & 50     & 12         & 231 & -- & --     & --     & --             & satellite, Marchis et al. (2014), Rozehnal and Bro\v{z} (2014)\\
005 & 3548 Eurybates & 50     & 218        & 68  & 87 & --     & CP  & 0.06        & Roig et al. (2008), Bro\v{z} and Rozehnal (2011) \\
006 & 9799 1996 RJ   & 60     & 7          & 72  & 26 & --     & --     & 0.06        & Rozehnal and Bro\v{z} (2014)\\ 
007 & James Bond     & $\infty$ & 1        & (himself)& -- & -- & ASP  & variable  & Campbell et al. (1995) \\   
008 & 20961 Arkesilaos & 50   & 37         & --  & -- & --     & --     & --            & Rozehnal and Bro\v{z} (2014)\\
009 & 4709 Ennomos   & 100    & 30         & (1867,4709) & --  &  --    & --     & 0.06        & Rozehnal and Bro\v{z} (2014)\\  
010 & 247341 2001 UV209 & 100 & 13         & --  & -- & --     & --     & 0.09        & Rozehnal and Bro\v{z} (2014)\\       

\multicolumn{10}{c}{\normalsize \it Inner Main Belt, $2.0  < a < 2.5$ AU, $i<17.5^\circ$}\\[1.mm]                      
401 & 4 Vesta        & 50     & 15252      & 525  & 50 & $1.5\pm0.5$     & V    &  0.35       & source of HEDs, two overlaping families? \\
402 & 8 Flora        & 60     & 13786      & (8,254) & -- & $2.5\pm0.5$    & S    &  0.30       & dispersed, source of LL NEAs, Dykhuis et al. (2014)\\ 
403 & 298 Baptistina & 45     & 2500       & 21   & -- & $0.25\pm0.05$   & X    &  0.16       & related to K/T impact? Bottke et al. (2007)\\ 
404 & 20 Massalia    & 55     & 6424       & 132  & 27 & $0.25\pm0.05$   & S    &  0.22       & Vokrouhlick\'y et al. (2006b)  \\
405 & 44 Nysa-Polana & 50     & 19073      & (135,142,495) & -- & $1.0\pm0.5$  & SFC  &  0.28/0.06  & Walsh et al. (2013), Dykhuis and Greenberg (2015)\\ 
406 & 163 Erigone    & 50     & 1776       & 72   & 46 & $0.2\pm0.05$    & CX  &  0.06        & Vokrouhlick\'y et al. (2006b)    \\    
407 & 302 Clarissa   & 55     & 179        & 34   & 15 & $0.05\pm0.01$   & X    &  0.05       & compact with ears, cratering \\
408 & 752 Sulamitis  & 55     & 303        & 61   & 35 & $0.3\pm0.1$     & C    &  0.04       &      \\ 
409 & 1892 Lucienne  & 100    & 142        & 11   & 11 & $0.15\pm0.05$   & S    &  0.22       &      \\
410 & 27 Euterpe     & 65     & 474        & 110  & 16 & $0.50\pm0.25$   & S    &  0.26       &      \\
411 & 1270 Datura    & 10     & 6          & 8    & 3  & --              & S    &  0.21       & Nesvorn\'y et al. (2006c)     \\
412 & 21509 Lucascavin & 10   & 3          & --   & -- & --              & S    &  --         & Nesvorn\'y and Vokrouhlick\'y (2006)    \\
413 & 84 Klio        & 130$^*$  & 330      & 78   & 33 & $0.75\pm0.25$   & C    &  0.07       & interloper 12?, Masiero et al. (2013)\\
414 & 623 Chimaera   & 120    & 108        & 43   & 21 & $0.3\pm0.1$     & CX   &  0.06       & Masiero et al. (2013)\\
415 & 313 Chaldaea   & 130$^*$  & 132      & (313,1715) & -- & $1.0\pm0.5$ & C  &  0.07       & 1715 in Masiero et al. (2013)\\
416 & 329 Svea       & 150    & 48         & 70   & 21 &  $0.3\pm0.1$    & CX    &  0.06      & new, near 3:1 \\
417 & 108138 2001 GB11 & 20   & 9          & --   & -- & --            & --   & --            & new, compact \\
\multicolumn{10}{c}{\normalsize \it Inner Main Belt, $2.0  < a < 2.5$ AU, $i>17^\circ$}\\[1.mm]     
701 & 25 Phocaea     & 150    & 1989       & (25,587) & -- & $2.0\pm1.0$   & S     & 0.22       & Carruba (2009b), Carruba et al. (2010) \\
\hline
\end{tabular}
\end{center}
\end{sidewaystable}

\clearpage
\setcounter{table}{1}
\begin{sidewaystable}[h!]
\small
\vspace*{10.cm}
\caption{Continued.}
\begin{center}  
\begin{tabular}{llrrrrrrrl}
\hline
 FIN    & Family  & $d_{\rm cut}$   & \# of & $D_{\rm LM}$ & $D_{\rm frag}$ & $C_0$        & Tax. & $p_{\rm V}$ & References and Notes     \\
       & Name    &  (m s$^{-1}$)   & mem.  & (km)        & (km)       & ($10^{-4}$ AU)   & Type &            &            \\    
\hline
\multicolumn{10}{c}{\normalsize \it Central Main Belt, $2.5  < a < 2.82$ AU, $i<17.5^\circ$}\\[1.mm]              
501 & 3 Juno         & 55     & 1684       & 231  & 25 & $0.5\pm0.2$          & S    & 0.25     & cratering, relation to H chondrites? \\
502 & 15 Eunomia     & 50     & 5670       & 256  & 100 & $2.0\pm0.7$          & S    & 0.19     & continues beyond 5:2?\\
503 & --             & --     & --         & --   & -- & --                   & --   & --       & 46 Hestia moved to candidates  \\
504 & 128 Nemesis    & 50     & 1302       & 178  & 50 & $0.25\pm0.05$        & C    & 0.05     & 3827 in Milani et al. (2014), 125 in Cellino et al. (2002)\\
505 & 145 Adeona     & 50     & 2236       & 141  & 78 & $0.7\pm0.3$          & C    & 0.07     &  \\
506 & 170 Maria      & 60     & 2940       & (472,170) & -- & $2.0\pm1.0$       & S    & 0.25     & (472) Roma in Masiero et al. (2013)\\
507 & 363 Padua      & 45     & 1087       & 91   & 48 & $0.5\pm0.2$          & X  & 0.10      & Carruba (2009a), also known as the (110) Lydia family\\
508 & 396 Aeolia     & 20     & 296        & 46   & 13 & $0.075\pm0.025$      & X  & 0.17       & compact, young? \\
509 & 410 Chloris    & 80     & 424        & 107  & 56 & $0.75\pm0.25$        & C    & 0.06     & eroded  \\
510 & 569 Misa       & 50     & 702        & 65   & 57  & $0.5\pm0.2$          & C    & 0.03     & V-shaped subfamily inside\\
511 & 606 Brangane   & 55     & 195        & 36   & 18 & $0.04\pm0.01$        & S    & 0.10     & compact, 606 offset, interloper? \\
512 & 668 Dora       & 45     & 1259       & (1734,668) & --  & --              & C    & 0.05     & 668 offset, 1734 in Masiero et al. (2013), V-shaped subfamily\\
513 & 808 Merxia     & 55     & 1215       & 34   & 28 & $0.3\pm0.1$          & S    & 0.23     & Vokrouhlick\'y et al. (2006b)   \\
514 & 847 Agnia      & 30     & 2125       & (847,3395) & --  & $0.15\pm0.05$   & S    & 0.18     & $z_1$ resonance, Vokrouhlick\'y et al. (2006c) \\
515 & 1128 Astrid    & 60     & 489        & 42   & 29 & $0.12\pm0.02$        & C    & 0.08     & Vokrouhlick\'y et al. (2006b)  \\
516 & 1272 Gefion    & 50     & 2547       & (2595,1272) & -- & $0.8\pm0.3$     & S    & 0.20     & source of L chondrites? Nesvorn\'y et al. (2009), also known as 93 and 2595\\
517 & 3815 Konig     & 55     & 354        & 22   & 34 & $0.06\pm0.03$        & CX    & 0.04     & compact, young? Nesvorn\'y et al. (2003), 342 and 1639 offset  \\
518 & 1644 Rafita    & 70     & 1295       & (1658,1587) & -- & $0.5\pm0.2$     & S    & 0.25     & 1644 probably interloper\\
519 & 1726 Hoffmeister & 45   & 1819       & (272,1726)  & -- & $0.20\pm0.05$   & CF  & 0.04     & (272) Antonia in Masiero et al. (2013), but 272 offset\\
520 & 4652 Iannini   & 25     & 150        & 5    & 10 & --                   & S    & 0.32     & 1547 offset, compact, Nesvorn\'y et al. (2003) \\
521 & 7353 Kazuya    & 50     & 44         & 11   & 10 & --                   & S    & 0.21     & small clump   \\
522 & 173 Ino        & 50     & 463        & 161  & 21 & $0.5\pm0.2$          & S    & 0.24     & also known as 18466, large and dark 173 is probably interloper, ears?    \\
523 & 14627 Emilkowalski & 10 & 4          & 7    & 3 & --                   & S    & 0.20     & Nesvorn\'y and Vokrouhlick\'y (2006) \\
524 & 16598 1992 YC2 & 10     & 3          & --   & --& --                   & S    & --       & Nesvorn\'y and Vokrouhlick\'y (2006) \\  
525 & 2384 Schulhof  & 10     & 6          & 12   & 4 & --                   & S    & 0.27     & Vokrouhlick\'y and Nesvorn\'y (2011)\\
526 & 53546 2000 BY6 & 40     & 58         & 8    & 18 & --                   & C    & 0.06     & Milani et al. (2014)    \\
527 & 5438 Lorre     & 10     & 2          & 30   & -- & --                   & C    & 0.05     & Novakovi\'c et al. (2012)    \\
528 & 2782 Leonidas  & 50     & 135        & (4793,2782) & --   & --            & CX   & 0.07     & new, related to 144?    \\
529 & 144 Vibilia    & 100$^*$   & 180     & 142  & -- & --                   & C    & 0.06     & Masiero et al. (2013), PDS list identical to 2782  \\
530 & 322 Phaeo      & 100$^*$   & 146     & 72   & 31 &  $0.3\pm0.1$         & X   & 0.06       & Cellino et al. (2002), joins (2669) Shostakovich \\
531 & 2262 Mitidika  & 100$^*$   & 653     & (404,5079) & -- & --             & C    & 0.06     & dispersed, 404 offset, 2262 has $p_{\rm V}=0.21$ \\
532 & 2085 Henan     & 50     & 1872       & 18   & 32 & $0.75\pm0.25$        & L    & 0.20     & 2085 offset in $i_{\rm P}$, 4 families in Milani et al. (2014) \\  
533 & 1668 Hanna     & 60     & 280        & 22   & 32 & $0.2\pm0.1$          & CX  & 0.05      & Masiero et al. (2013)  \\
534 & 3811 Karma     & 60     & 124        & 26   & 24 & $0.25\pm0.05$        & CX  & 0.05      & Milani et al. (2014)   \\
\hline
\end{tabular}
\end{center}
\end{sidewaystable}

\clearpage
\setcounter{table}{1}
\begin{sidewaystable}[h!]
\small
\vspace*{10.cm}
\caption{Continued.}
\begin{center}  
\begin{tabular}{llrrrrrrrl}
\hline
  FIN    & Family  & $d_{\rm cut}$   & \# of & $D_{\rm LM}$   & $D_{\rm frag}$  & $C_0$        & Tax. & $p_{\rm V}$ & References and Notes     \\
       & Name    &  (m s$^{-1}$)    & mem.  & (km)   & (km) & ($10^{-4}$ AU)    & Type &            &            \\            
\hline
\multicolumn{10}{c}{\normalsize \it Central Main Belt, $2.5  < a < 2.82$ AU, $i<17.5^\circ$}\\[1.mm]      
535 & 2732 Witt      & 45     & 1816     & 11   & 25 & $0.75\pm0.25$          & S    & 0.26     & relation to the Charis family beyond 5:2? 10955 and 19466 in Milani et al.    \\  
536 & 2344 Xizang    & 65     & 275      & 17   & 20 & $0.3\pm0.1$            & --    & 0.12     & Milani et al. (2014), includes 396  \\ 
537 & 729 Watsonia   & 130    & 99       & 52   & 38 & --                     & L    & 0.13     & Cellino et al. (2014)   \\  
538 & 3152 Jones     & 40     & 22       & 37   & 11 & --                     & T    & 0.05     & new, compact, diagonal in $(a_{\rm P},e_{\rm P})$  \\  
539 & 369 Aeria      & 90     & 272      & 75   & 17 & $0.3\pm0.1$            & X   & 0.17     & new, part above 2.6778 AU down in $i_{\rm P}$   \\  
540 & 89 Julia       & 70     & 33       & 147  & 6 & --                     & S    & 0.19     & new, compact   \\  
541 & 1484 Postrema  & 100    & 108      & 47   & -- & --                     & CX   & 0.05     & new, large 387,547,599?  \\  

\multicolumn{10}{c}{\normalsize \it Central Main Belt, $2.5  < a < 2.82$ AU, $i>17.5^\circ$}\\[1.mm] 
801 & 2 Pallas       & 350    & 128      & 513  & 40 & --                  & B     & 0.16      & Carruba et al. (2010, 2012), part beyond 5:2   \\     
802 & 148 Gallia     & 200    & 182      & 81   & 19 & $0.5\pm0.1$         & S     & 0.17      & large interlopers\\  
803 & 480 Hansa      & 200    & 1094     & 56   & 62 & --                  & S     & 0.26      & 2 families in Milani et al. (2014)   \\  
804 & 686 Gersuind   & 120    & 415      & 49   & 36 & --                  & S     & 0.15      & 2 families in Milani et al. (2014) \\
805 & 945 Barcelona  & 150    & 306      & 27   & 19 & $0.25\pm0.05$       & S     & 0.25      & 2 families in Milani et al. (2014) \\
806 & 1222 Tina      & 200    & 96       & 26   & 10 & $0.10\pm0.05$       & X     & 0.34      & in the $g-g_6=0$ resonance, Carruba and Morbidelli (2011)\\  
807 & 4203 Brucato   & 200    & 342      & 18   & 44 & $0.5\pm0.2$         & CX     & 0.06      & 1263 interloper? Carruba (2010), 4 families in Milani et al. (2014)   \\ 
\multicolumn{10}{c}{\normalsize \it Outer Main Belt, $2.82 < a < 3.7$ AU, $i<17^\circ$} \\[1.mm]    
601   & 10 Hygiea      & 60     & 4854   & 428  & -- & --                  & CB  & 0.06      & Carruba et al. (2014)   \\ 
602   & 24 Themis      & 60     & 4782   & 177  & 230 & $2.5\pm1.0$         & C    & 0.07      & includes 656 Beagle, Nesvorn\'y et al. (2008b)  \\ 
603   & 87 Sylvia      & 130    & 255    & 263  & -- & --                  & X  & 0.05      & Vokrouhlick\'y et al. (2010)   \\ 
604   & 137 Meliboea   & 85     & 444    & (511,137) & -- & --               & C    & 0.05     &  (511) Davida in Masiero et al. (2013)  \\    
605   & 158 Koronis    & 45     & 5949   & (208,158,462) & -- & $2.0\pm1.0$  & S    & 0.15     &  (208) Lacrimosa in Masiero et al. (2013)  \\ 
606   & 221 Eos        & 45     & 9789   & (221,579,639) & -- & $1.5\pm0.5$  & K    & 0.13      & Vokrouhlick\'y et al. (2006a), Bro\v{z} and Morbidelli (2013) \\ 
607   & 283 Emma       & 40     & 76     & 122  & 56 & $0.3\pm0.1$         & C   & 0.05      &  affected by the z$_1$ resonance?  \\ 
608   & 293 Brasilia   & 50     & 579    & (3985) & -- & $0.20\pm0.05$       & X    & 0.18      &  293 interloper?, also known as (1521) Sejnajoki, Nesvorn\'y et al. (2003)\\ 
609   & 490 Veritas    & 30     & 1294   & 113  & 78 & $0.2\pm0.1$         & CPD  & 0.07      &  see Section 3  \\ 
610   & 832 Karin      & 10     & 541    & 17   & 16 & $0.03\pm0.01$       & S    & 0.21     &   see Section 3, Harris et al. (2009) \\ 
611   & 845 Naema      & 35     & 301    & 61   & 37 & $0.20\pm0.05$       & C    & 0.08      &    \\ 
612   & 1400 Tirela    & 50     & 1395   & (1040,1400) & -- & $0.75\pm0.25$  & S    & 0.07      & 8 families in Milani et al. (2014)   \\ 
613   & 3556 Lixiaohua & 45     & 756    & (3330,3556) & -- & $0.25\pm0.05$  & CX  & 0.04      & 3330 offset, Novakovi\'c et al. (2010)   \\ 
614   & 9506 Telramund & 45     & 468    & (179,9506)  & -- & --             & S    & 0.22      & (179) Klytaemnestra in Masiero et al. and Milani et al.   \\ 
615   & 18405 FY12     & 50     & 104    & 9           & 14 & $0.08\pm0.03$       & CX  & 0.17      &    \\ 
\hline
\end{tabular}
\end{center}
\end{sidewaystable}

\clearpage
\setcounter{table}{1}
\begin{sidewaystable}[h!]
\small
\vspace*{10.cm}
\caption{Continued.}
\begin{center}  
\begin{tabular}{llrrrrrrrl}
\hline
FIN    & Family  & $d_{\rm cut}$   & \# of & $D_{\rm LM}$   & $D_{\rm frag}$  & $C_0$        & Tax. & $p_{\rm V}$ & References and Notes     \\
       & Name    &  (m s$^{-1}$)  & mem.  & (km)          & (km) & ($10^{-4}$ AU)   & Type &            &            \\    
\hline
\multicolumn{10}{c}{\normalsize \it Outer Main Belt, $2.82 < a < 3.7$ AU, $i<17.5^\circ$} \\[1.mm]    
616   & 627 Charis     & 80     & 808  & 50   & 45 & --                & C    & 0.08      & 16286 in Milani et al. (2014), related to Witt family?  \\       
617   & 778 Theobalda  & 60     & 376  & 66   & 50 & --                & CX    & 0.06      & 778 offset, Novakovi\'c et al. (2010)   \\       
618   & 1189 Terentia  & 60     & 79   & 63   & 18 & $0.13\pm0.03$     & C    & 0.07      &    \\       
619   & 10811 Lau      & 120    & 56   &  8   & 6 & $0.075\pm0.025$   & S    & 0.27      & 36824 interloper?   \\       
620   & 656 Beagle     & 25     & 148  & 54   & 28 & $0.07\pm0.03$     & C    & 0.09     & 656 and 90 offset, Nesvorn\'y et al. (2008b) \\       
621   & 158 Koronis(2) & 8      & 246  & 34   & 13 & $0.010\pm0.005$   & S    & 0.14     & young, Molnar and Haegert (2009) \\       
622   & 81 Terpsichore & 120    & 138  & 123  & 27 & $0.50\pm0.25$     & C    & 0.05     &    \\       
623   & 709 Fringilla  & 150    & 134  & 96   & 55 & --                & X    & 0.05     & large 1191   \\       
624   & 5567 Durisen   & 100    & 27   & 34   & 23 & --                & X    & 0.04     &    \\       
625   & 5614 Yakovlev  & 120    & 67   & 13   & 23 & $0.15\pm0.05$     & C    & 0.05     &    \\       
626   & 7481 San Marcello & 100 & 144  & (3978,7489) & --      & --                & X    & 0.19     & also known as 12573  \\       
627   & 15454 1998 YB3    & 80  & 38   & (3156,15454) & --   & $0.10\pm0.05$     & CX    & 0.05     & \\       
628   & 15477 1999 CG1 & 95     & 248  & --   & -- & --               & S    & 0.10     &    \\       
629   & 36256 1999 XT17 & 70    & 58   & (15610,36256) & --     & $0.25\pm0.05$     & S    & 0.21     & several large bodies   \\       
630   & 96 Aegle        & 50    & 99   & 165  & 38 & --                & CX  & 0.07      & Masiero et al. and Milani et al.  \\       
631   & 375 Ursula      & 70    & 1466 & (1306,375) & -- & --            & CX    & 0.06     &  Masiero et al. and Milani et al.  \\       
632   & 618 Elfriede    & 40    & 63   & 122  & 26 & --                & C    & 0.05     & compact, recent?   \\       
633   & 918 Itha        & 100   & 54   & 21   & 35 & --                & S    & 0.23     & dispersed, many sizable members   \\       
634   & 3438 Inarradas  & 80    & 38   & 25   & 33 & --                & CX  & 0.07      & Milani et al. (2014)   \\       
635   & 7468 Anfimov    & 60    & 58   & 10   & 14 & --                & S    & 0.16     & Milani et al. (2014)   \\       
636   & 1332 Marconia   & 30    & 34   & 50   & 16 & --                & CX     & 0.05   & new   \\       
637   & 106302 2000 UJ87 & 60   & 64   & 7    & 15 & --                & CX     & 0.05   & new, large 132999   \\       
638   & 589 Croatia     & 45    & 93   & 92   & 31 & $0.5\pm0.2$       & X     & 0.07   & new, 21885 in Milani et al. (2014) \\       
639   & 926 Imhilde     & 70    & 43   & 50   & 18 & $0.2\pm0.1$       & CX     & 0.05   & new   \\       
640   & P/2012 F5 (Gibbs) & 10  & 8    & --   & -- & --                & --    & --      & Novakovi\'c et al. (2014)   \\       
641   & 816 Juliana     & 80    & 76   & 68   & 39 & --                & CX    & 0.05    & Masiero et al. (2013)    \\       
\multicolumn{10}{c}{\normalsize \it Outer Main Belt, $2.82 < a < 3.5$ AU, $i>17.5^\circ$} \\[1.mm]  
901   & 31 Euphrosyne   & 120   & 2035  & 276  & 130 & --               & C    & 0.06     &  Carruba et al. (2014), 3 families in Milani et al. (2014)   \\      
902   & 702 Alauda      & 120   & 1294  & 191  & -- & $2.5\pm1.0$   & B    & 0.07        & 276 and 1901 offset, 4 families in Milani et al. (2014)   \\      
903   & 909 Ulla        & 120   & 26    & 113  & 28 & --               & X  & 0.05      &    \\      
904   & 1303 Luthera    & 50    & 163   & 87   & 56 & --               & X    & 0.04     & also known as (781) Kartivelia   \\      
905   & 780 Armenia     & 50    & 40    & 98   & 22 & --               & C  & 0.05      &  compact  \\             
\hline
\end{tabular}
\end{center}
\end{sidewaystable}

\end{document}